\documentclass[journal=tches,eprint]{iacrtrans}
\usepackage[T1]{fontenc}
\usepackage{graphicx}
\usepackage{amsmath}
\usepackage{mathtools}
\usepackage{amsfonts}
\usepackage{xargs}
\usepackage[breaklinks,colorlinks=true, citecolor=blue]{hyperref}
\usepackage[ruled,vlined,linesnumbered]{algorithm2e}
\DontPrintSemicolon
\usepackage{cleveref}
\usepackage[colorinlistoftodos,prependcaption,textsize=tiny]{todonotes}
\usepackage{xcolor}
\usepackage{listings}
\usepackage{caption}
\usepackage{booktabs}
\usepackage{tikz-cd}
\usepackage{siunitx}
\usepackage{float}
\usepackage{comment}
\usepackage{multirow}

\definecolor{codegreen}{rgb}{0,0.6,0}
\definecolor{codegray}{rgb}{0.5,0.5,0.5}
\definecolor{codepurple}{rgb}{0.58,0,0.82}
\definecolor{backcolour}{rgb}{0.95,0.95,0.92}

\lstdefinestyle{overleafDocDefault}{
    backgroundcolor=\color{backcolour},
    commentstyle=\color{codegreen},
    keywordstyle=\color{magenta},
    numberstyle=\tiny\color{codegray},
    stringstyle=\color{codepurple},
    basicstyle=\ttfamily\footnotesize,
    breakatwhitespace=false,
    breaklines=true,
    captionpos=b,
    keepspaces=true,
    numbers=left,
    numbersep=5pt,
    showspaces=false,
    showstringspaces=false,
    showtabs=false,
    tabsize=2
}
\lstdefinestyle{cleanC}{
  language=C,
  backgroundcolor=\color{white},
  frame=single,
  rulecolor=\color{black!30},
  framesep=8pt,
  xleftmargin=12pt,
  xrightmargin=12pt,
  basicstyle=\ttfamily\small,
  keywordstyle=\bfseries\color{blue!70!black},
  commentstyle=\itshape\color{green!50!black},
  stringstyle=\color{red!60!black},
  numbers=left,
  numberstyle=\tiny\color{gray},
  numbersep=8pt,
  breaklines=true,
  tabsize=4,
  showstringspaces=false,
  columns=flexible,
  aboveskip=1em,
  belowskip=1em,
  abovecaptionskip=0.5em,
}
\lstdefinestyle{smallC}{
  language=C,
  backgroundcolor=\color{white},
  basicstyle=\scriptsize\ttfamily,
  keywordstyle=\bfseries\color{blue!55!black},
  commentstyle=\itshape\color{gray!70!black},
  stringstyle=\color{red!50!black},
  numbers=left,
  numberstyle=\tiny\color{gray},
  numbersep=8pt,
  stepnumber=1,
  frame=single,
  rulecolor=\color{gray!35},
  frameround=tttt,
  xleftmargin=2.4em,
  framexleftmargin=2.4em,
  escapeinside={(*@}{@*)},
  columns=fullflexible,
  keepspaces=true,
  showstringspaces=false,
  breaklines=true,
  tabsize=4,
  captionpos=b,
    literate=
    {_}{{\_}}1
    {^}{{\textasciicircum}}1
    {&}{{\&}}1
}

\lstdefinestyle{smallARM}{
  language=[x86masm]Assembler, 
  backgroundcolor=\color{white},
  basicstyle=\scriptsize\ttfamily,
  keywordstyle=\bfseries\color{blue!55!black}, 
  commentstyle=\itshape\color{gray!70!black}, 
  stringstyle=\color{red!50!black}, 
  numbers=left,
  numberstyle=\tiny\color{gray},
  numbersep=8pt,
  stepnumber=1,
  frame=single,
  rulecolor=\color{gray!35},
  frameround=tttt,
  xleftmargin=2.4em,
  framexleftmargin=2.4em,
  escapeinside={(*@}{@*)},
  columns=fullflexible,
  keepspaces=true,
  showstringspaces=false,
  breaklines=true,
  tabsize=4,
  captionpos=b,
  literate=
    {_}{{\_}}1
    {^}{{\textasciicircum}}1
    {&}{{\&}}1,
  morekeywords={ 
    MOV, MOVS, MOVW, MOVT,
    ADD, ADDS, ADC, ADR,
    SUB, SUBS, SBC, RSB,
    MUL, MULS, MLS,
    AND, ORR, EOR, BIC,
    LSL, LSR, ASR, ROR,
    CMP, CMN, TST, TEQ,
    B, BL, BX, BLX,
    PUSH, POP,
    LDR, STR, LDRB, STRB,
    LDRH, STRH, LDRSH, LDRSB,
    NOP, SEV, WFE, WFI,
    CPS, MRS, MSR,
    SVR, DSPS, ISB, DMB, DSB,
    IT, NEG, NOT,
    ADDW, SUBW, MOVT, MOVW,
    CBZ, CBNZ, TBB, TBH,
    R0, R1, R2, R3, R4, R5, R6, R7,
    R8, R9, R10, R11, R12, SP, LR, PC,
    APSR, IPSR, EPSR, IEPSR, MSP, PSP, PRIMASK, BASEPRI, FAULTMASK, CONTROL,
    .text, .data, .bss, .global, .thumb, .thumb_func,
    .align, .word, .byte, .space, .equ, .end
  },
  alsoletter={R}, 
  alsoother={.}, 
}

\definecolor{yhl}{rgb}{0.98, 0.95, 0.75}
\definecolor{rhl}{rgb}{1.0, 0.85, 0.90}
\definecolor{bhl}{rgb}{0.85, 0.90, 1.0}

\newcommand{\highlight}[2]{\colorbox{#1}{\strut #2}}

\newcommand{\hw}{\ensuremath{\mathsf{hw}}\xspace}

\newcommand{\hqc}{\textsc{HQC}\xspace}

\newcommand{\ct}{\ensuremath{\mathsf{ct}}\xspace}

\newcommand{\Ring}{\ensuremath{\mathcal{R}}\xspace}
\newcommand{\Rot}{\ensuremath{\operatorname{rot}}\xspace}

\newcommand{\fafft}{\ensuremath{\mathsf{FAFFT}}\xspace}

\SetKwData{PK}{PK}
\SetKwData{SK}{SK}
\newcommand{\sk}{\SK}
\newcommand{\pk}{\PK}

\newcommand{\F}{\mathbb{F}}
\newcommand{\FF}{\F}

\newcommand{\vmul}{\texttt{vect\_mul}\xspace}
\newcommand{\vadd}{\texttt{vect\_add}\xspace}
\newcommand{\kmul}{\texttt{karatsuba\_mul}\xspace}
\newcommand{\smul}{\texttt{schoolbook\_mul}\xspace}
\newcommand{\vsup}{\texttt{vect\_write\_support\_to\_vector}\xspace}
\newcommand{\refC}{the reference C implementation of HQC\xspace}
\newcommand{\ssv}{secret sparse vector\xspace}
\newcommand{\ssvs}{secret sparse vectors\xspace}

\newcommand{\uint}{\texttt{uint64\_t}\xspace}

\renewcommand{\vec}[1]{\ensuremath{\mathbf{#1}}\xspace}
\DeclareMathOperator{\supp}{supp}

\newcommandx{\unsure}[2][1=]{\todo[linecolor=red,backgroundcolor=red!25,bordercolor=red,#1]{#2}}
\newcommandx{\change}[2][1=]{\todo[linecolor=blue,backgroundcolor=blue!25,bordercolor=blue,#1]{#2}}
\newcommandx{\info}[2][1=]{\todo[linecolor=green,backgroundcolor=green!25,bordercolor=green,#1]{#2}}
\newcommandx{\improvement}[2][1=]{\todo[linecolor=purple,backgroundcolor=purple!25,bordercolor=purple,#1]{#2}}

\SetKwProg{Function}{function}{}{}
\SetKwFunction{SHATHREE}{SHA3-512}
\SetKwFunction{PKEKeyGen}{HQC-PKE.KeyGen}
\SetKwFunction{PKEEncrypt}{HQC-PKE.Encrypt}
\SetKwFunction{PKEDecrypt}{HQC-PKE.Decrypt}
\SetKwFunction{KEMKeyGen}{HQC-KEM.KeyGen}
\SetKwFunction{KEMEncaps}{HQC-KEM.Encaps}
\SetKwFunction{KEMDecaps}{HQC-KEM.Decaps}
\SetKwFunction{SampleUniform}{SampleUniform}
\SetKwFunction{SampleVect}{SampleVect}
\SetKwFunction{SampleFixedWeightVect}{SampleFixedWeightVect}
\SetKwFunction{Encode}{Encode}
\SetKwFunction{Decode}{Decode}
\SetKwFunction{Hash}{Hash}
\SetKwFunction{XOFInit}{XOF.Init}
\SetKwFunction{XOFGetBytes}{XOF.GetBytes}
\SetKwFunction{Truncate}{Truncate}
\SetKwFunction{PRG}{PRG}
\SetKwFunction{KDF}{KDF}
\SetKwFunction{Valid}{Valid}
\SetKwData{salt}{salt}
\SetKwData{seed}{seed}
\SetKwData{seedPKE}{seedPKE}
\SetKwData{seedPKEek}{seedPKE.ek}
\SetKwData{seedPKEdk}{seedPKE.dk}
\SetKwData{seedKEM}{seedKEM}
\SetKwData{ekPKE}{ekPKE}
\SetKwData{dkPKE}{dkPKE}
\SetKwData{ekKEM}{ekKEM}
\SetKwData{dkKEM}{dkKEM}
\SetKwData{cPKE}{cPKE}
\SetKwData{cKEM}{cKEM}
\SetKwData{ctx}{ctx}
\SetKwData{ctxPKEdk}{ctxPKE.dk}
\SetKwData{ctxPKEek}{ctxPKE.ek}
\SetKwData{ctxKEM}{ctxKEM}

\newcommand{\nnz}{62}        
\newcommand{\zerofrac}{88.8} 

\begin{document}

\title{Exploiting Load/Store Leakage of Sparse Vectors for Key Recovery in HQC}
\author{Gustavo Banegas\inst{1} \and Benjamin Smith\inst{1} \and Jad Zahreddine\inst{1,2}}
\authorrunning{Gustavo Banegas, Benjamin Smith, Jad Zahreddine}
\institute{
       LIX, Inria, CNRS, École Polytechnique, Institut Polytechnique de Paris, France
 \email{gustavo@cryptme.in, smith@lix.polytechnique.fr}
 \and
  eShard, France  \\
  \email{jad.zahreddine@eshard.com}
}

\maketitle              
\begin{abstract}
    Hamming Quasi-Cyclic (HQC) is a code-based key encapsulation mechanism
    selected by NIST for standardization,
    making its resistance to implementation attacks critically important.
    We present a side-channel attack that exploits load/store leakage
    in the manipulation of HQC's sparse secret vectors.
    Analysing Cortex-M4 assembly generated from the reference
    implementation, we identify a leakage surface in which the low and
    high 32-bit halves of each 64-bit word leak with different strengths,
    due to compiler-generated register spilling.
    We exploit this leakage to construct a simple zero-word distinguisher
    classifying machine words of the secret vector as zero or nonzero
    from electromagnetic measurements.
    The recovered zero positions are then translated into decoding hints,
    reducing HQC key recovery to a shortened syndrome-decoding problem.
    We analyse the resulting decoding complexity for all HQC parameter sets:
    at 32-bit granularity an expected $88.7\%$ of the machine words of~$y$ are
    zero for HQC-1, cutting the decoding to ${\approx}\,2^{46}$ bit operations.
    Experiments on a Cortex-M4 validate the predicted low/high-half
    asymmetry---approximately $500$ traces for the stronger low-half channel
    and $5{,}000$ for the weaker high-half channel---and recover
    the zero words of an HQC-1 key at 32-bit granularity.
    Finally, we discuss practical countermeasures that eliminate
    the sparsity exploited by the attack.

\end{abstract}

\begingroup
\makeatletter
\def\@thefnmark{$\*$}\relax
\@footnotetext{\relax
Author list in alphabetical order; see
\url{https://ams.org/profession/leaders/CultureStatement04.pdf}.

This work was supported by:
\begin{itemize}
  \item the HYPERFORM consortium, funded by France
  through Bpifrance, and by the France 2030 program under
  grant agreement ANR-22-PETQ-0008 PQ-TLS.
\item the SPARTAQUS consortium, funded by Germany through
  Agentur für Innovation in der Cybersicherheit GmbH.
\end{itemize}

\def\ymdtoday{\leavevmode\hbox{\the\year-\twodigits\month-\twodigits\day}}
\def\twodigits#1{\ifnum#1<10 0\fi\the#1}

Date of this document: \ymdtoday.
}
\endgroup

\section{Introduction}
\label{sec:intro}
Hamming Quasi-Cyclic (\hqc~\cite{gaborit2025hamming}) has been selected by
NIST for standardization as a post-quantum key encapsulation mechanism
(KEM)~\cite{nistpqc}, as a code-based complement for the lattice-based
standard ML-KEM~\cite{FIPS203}.
As \hqc moves toward standardization and deployment,
its resistance to side-channel attacks---not only its algorithmic
security---must be scrutinised.

Simple and differential power analysis (SPA/DPA) are well-established threats.
The central questions are \emph{which} operations leak exploitable information
and \emph{how} that leakage can aid key recovery.
Prior work has mostly targeted \emph{arithmetic} operations such as polynomial
multiplication~\cite{BockBBCPS24,YangRZSB23} or other secret-dependent
intermediates~\cite{RaviPJDB24}; leakage from merely \emph{loading} or
\emph{storing} secret data has drawn little attention.
For lattice-based schemes this is unsurprising:
secret coefficients come from a non-trivial distribution on small
integers,
so the Hamming weight of a loaded machine word reveals little
information on these coefficients.

\hqc is different: its secret vector~$y$ is binary, and extremely sparse
($\omega = 66$ ones among $n = 17\,669$ bits for \hqc-1), so the overwhelming
majority of the machine words encoding~$y$ are identically zero. Marshall,
Page, and Webb~\cite{miracle} showed that on ARM Cortex-M4, consecutive load and
store instructions leak the Hamming weight---and the Hamming distance---of the
data on the memory bus, even across register-disjoint instructions. A single
trace can therefore separate a zero word from a nonzero one at each step of the
encoding loop. We turn this into a key-recovery attack by casting the problem as
\emph{decoding with hints}: every distinguishable nonzero word localizes at least
one support position of~$y$, shrinking the search space.

\paragraph{Decoding with hints.}
A growing line of work incorporates partial information (\emph{hints}) about a
secret error vector into Information-Set Decoding (ISD) to lower the cost of the
underlying syndrome decoding problem (SDP).
The general framework in~\cite{DHorlemannPRSW21} covers several hint types
(known error or error-free locations,
partial message knowledge,
and sub-block weights),
transforming decoding instances into strictly smaller ones and
quantifying how many hints push the ISD work factor below the claimed security
level of Classic McEliece~\cite{mceliece1978public}, BIKE~\cite{BIKE}, and \hqc.
This is extended in~\cite{DAchilleEK26}
to form reliability vectors from
\emph{perfect} and \emph{approximate} hints, recasting the SDP as soft-decision
decoding that interpolates smoothly between ISD complexity and polynomial time.
Both works treat \emph{hint acquisition as a black box}. We close that gap: our
zero-word distinguisher produces, from electromagnetic traces, exactly the
known-zero-position hints of~\cite{DHorlemannPRSW21} and the reliability
information of~\cite{DAchilleEK26}, so the resulting instance feeds both methods; in this work we implement and
benchmark the ISD route and outline the statistical-decoding route.

\paragraph{Related work.}
Velek, Rabas, and Bucek~\cite{velek2026simplepoweranalysispolynomial} give a
single-trace SPA on the \emph{Additional} \hqc implementation, exploiting a
lookup-table base-case multiplication to recover secret bits four at a time. Goy,
Maillard, Gaborit, and Loiseau~\cite{GoyMGL24} mount a SASCA via belief
propagation on the Reed--Solomon codec to recover the \emph{shared key} from a
single electromagnetic trace.
Our work differs from these both in target and in generality:
unlike \cite{GoyMGL24}, we recover the
\emph{long-term private key}~$y$---reused across all
decapsulations---rather than a per-session key, and we exploit
load/store leakage of the sparse vector to feed a decoder rather than
read secret bits directly, so our hints plug
into~\cite{DHorlemannPRSW21} and~\cite{DAchilleEK26}.


\paragraph{Contributions.}
We present a practical load/store side-channel attack on \hqc that exploits the
leakage of the \ssv during polynomial multiplication, and carry
it through to full key recovery. Concretely:
\begin{itemize}
    \item \textbf{A new leakage surface.} We identify a previously unexplored
    load/store leakage in the \smul routine of the reference
        \hqc implementation (called by the higher-level Karatsuba
        multiplication), and show that the compiler-generated Cortex-M4 assembly
    (\texttt{-Os}) leaks each 64-bit word \emph{asymmetrically}: the low 32-bit
    half is both loaded and spilled to the stack, whereas the high half is only
    loaded, so the two halves leak at different rates.

    \item \textbf{A zero-word distinguisher.} We build a simple Welch's
        $t$-distinguisher that classifies each 32-bit half-word of~$y$
        as zero or nonzero.
        On an STM32F446RE it separates the low half in ${\approx}\,500$
        traces and the weaker high half in ${\approx}\,5\,000$,
        and all words are classified in parallel from a single batch of traces.

    \item \textbf{Hints from leakage.} We reduce key recovery to a
    \emph{shortened} syndrome-decoding instance whose known-zero hints feed
    directly into the decoding-with-hints frameworks
        of~\cite{DHorlemannPRSW21} and~\cite{DAchilleEK26}.
        At 32-bit (resp.~64-bit) granularity, an expected $88.7\%$
        (resp.~$78.7\%$) of the words of~$y$ are zero for \hqc-1.

    \item \textbf{Complexity of key recovery.} Exploiting the
    identity block of the shortened instance, we give a block-structured ISD that
    recovers the key, and we analyse its cost for all parameter sets. The
    half-word (32-bit) hints cut \hqc-1 decoding to ${\approx}2^{46}$ bit
    operations---about $40$\,s on a $48$-core machine---versus several days for
    word (64-bit) hints; the compiler induced lower bit granularity is thus what
    turns an impractical decoding into a routine one.

    \item \textbf{Countermeasures.} We describe two mitigations---per-call
    additive masking, and storing~$y$ in the FAFFT transform domain---that
    remove the sparsity the attack relies on, and we discuss their cost.
\end{itemize}

\section{Preliminary}
\label{sec:prem}
\subsection{Notation}
As usual, $\F_2$ denotes the finite field with two elements.
Given an integer $n \geq 1$,
we define the ring
\[
    \Ring
    :=
    \F_2[X]/(X^n - 1)
    \,.
\]
We identify polynomials in \Ring
with their coefficient vectors (with respect to the monomial basis)
in $\F_2^n$.
In this representation,
multiplication by some $a \in \Ring$ acts on vectors in $\F_2^n$
as multiplication by a circulant matrix $\Rot(a)$.
The Hamming weight of a vector $v \in \F_2^n$ is denoted by $\hw(v)$.

\subsection{\hqc in a nutshell}
\label{subsec:hqc-nutshell}

At a high level,
\hqc key generation (Algorithm~\ref{alg:HQC-PKE.KeyGen}) samples a public key $(h,s)\in \Ring^2$
and derives \ssvs $x,y\in\F_2^n$ from a short $\seed$
such that
\begin{equation}
    \label{eq:original-relation}
    s = x + h\cdot y \in \Ring
\end{equation}
(the product \(\cdot\) here is multiplication in \Ring,
not a vector dot product). The secret key $\sk$ can be the vectors $x,y$ or 
the $\seed$.  From Equation~\ref{eq:original-relation}, the public key $\pk$ is 
composed of a seed $r_h$ that generates $h$ and the result of the equation, namely, 
variable $s$. The $\mathsf{PRG}$ function is written as a simplification of 
the initialization of a hash function with the correct domain separator specified in the 
documentation of \hqc~\cite{gaborit2025hamming}.


Encapsulation (Algorithm~\ref{alg:HQC-KEM.Encaps}) uses fresh randomness
to form a ciphertext $\ct=(u,v)$,
while decapsulation (Algorithm~\ref{alg:HQC-KEM.Decaps}) reconstructs the
secret-dependent vector $y$ from $\seed$ and computes expressions
involving $u\cdot y$ (Algorithm~\ref{alg:HQC-PKE.Decrypt}).
This step is where our leakage model applies:
the multiplication routine loads the words of~$y$ in a regular pattern.

Here $n$ is the ambient vector length (the smallest primitive prime
greater than $n_1 n_2$), $k = 128$, $192$, or $256$ bits is the
shared-key size, $\omega$ is the fixed Hamming weight of the secret
vectors $(x,y)$, and $\omega_r = \omega_e$ is the fixed weight of the
ephemeral vectors $(r_1, r_2, e)$; all values are given in
Table~\ref{tab:parameters}.

In the following \(G\), \(H\), and \(I\) are SHA3-512 with domain
separation.

\noindent
\begin{minipage}[t]{0.48\textwidth}
\begin{algorithm}[H]
    \caption{\hqc PKE key generation.}
    \label{alg:HQC-PKE.KeyGen}
    \scriptsize
    \Function{\PKEKeyGen{\seed}}{
        \KwOut{Encryption key \ekPKE, decryption key \dkPKE}
        \(
            (\text{\seedPKEdk},\text{\seedPKEek})
            \gets
            I(\seedPKE)
        \)
        \;
        \(\ctxPKEdk \gets \text{\XOFInit{\seedPKEdk}}\)
        \;
        \(
            (\ctxPKEdk, \vec{y})
            \gets
            \text{\SampleFixedWeightVect{\ctxPKEdk, $\Ring_\omega$}} \label{line:leakage_key}
        \)
        \;
        \(
            (\ctxPKEdk, \vec{x})
            \gets
            \text{\SampleFixedWeightVect{\ctxPKEdk, $\Ring_\omega$}}
        \)
        \;
        \(\dkPKE \gets \seedPKEdk\)
        \;
        \(\ctxPKEek \gets \text{\XOFInit{\seedPKEek}}\)
        \;
        \(
            (\ctxPKEek,\vec{h})
            \gets
            \text{\SampleVect{\ctxPKEek,\Ring}}
        \)
        \;
        \( \vec{s} \gets \vec{x} + \vec{h}\cdot\vec{y} \)
        \;
        \(
            \ekPKE \gets (\seedPKEek, \vec{s})
        \)
        \;
        \Return{\((\ekPKE,\dkPKE)\)}
        \;
    }
\end{algorithm}
\end{minipage}\hfill
\begin{minipage}[t]{0.48\textwidth}
\begin{algorithm}[H]
    \caption{\hqc KEM key generation}
    \label{alg:HQC-KEM.KeyGen}
    \scriptsize
    \Function{\KEMKeyGen{}}{
        \(\seedKEM \stackrel{\$}{\gets} [0,256)^{32}\)
        \tcp*{32 random bytes}
        \(\ctxKEM \gets \text{\XOFInit{\seedKEM}}\)
        \;
        \(
            (\ctxKEM,\seedPKE)
            \gets
            \text{\XOFGetBytes{\ctxKEM,32}}
        \)
        \;
        \(
            (\ctxKEM,\sigma)
            \gets
            \text{\XOFGetBytes{\ctxKEM,32}}
        \)
        \;
        \(
            (\ekPKE,\dkPKE)
            \gets
            \text{\PKEKeyGen{\seedPKE}}
        \)
        \;
        \(\ekKEM \gets \ekPKE\)
        \;
        \(\dkKEM \gets (\ekKEM, \dkPKE, \sigma, \seedKEM)\)
        \;
        \Return{\((\ekKEM,\dkKEM)\)}
        \;
    }
\end{algorithm}
\end{minipage}

\noindent
\begin{minipage}[t]{0.48\textwidth}
\begin{algorithm}[H]
    \caption{\hqc PKE Encryption}
    \label{alg:HQC-PKE.Encrypt}
    \scriptsize
    \Function{\PKEEncrypt{\ekPKE,$\vec{m}$,$\theta$}}{
        \KwIn{
            Encryption key \ekPKE, message \vec{m}, randomness $\theta$
        }
        \KwOut{Ciphertext \((\vec{u},\vec{v})\)}
        \((\seedPKEek,\vec{s}) \gets \ekPKE\)
        \;
        \(\ctxPKEek \gets \text{\XOFInit{\seedPKEek}}\)
        \;
        \(
            (\ctxPKEek,\vec{h})
            \gets
            \text{\SampleVect{\ctxPKEek,\Ring}}
        \)
        \;
        \(\ctx_\theta \gets \text{\XOFInit{$\theta$}}\)
        \;
        \(
            (\ctx_\theta, \vec{r}_2)
            \gets
            \text{\SampleFixedWeightVect{$\ctx_\theta$, $\Ring_{\omega_r}$}}
        \)
        \;
        \(
            (\ctx_\theta, \vec{e})
            \gets
            \text{\SampleFixedWeightVect{$\ctx_\theta$, $\Ring_{\omega_e}$}}
        \)
        \;
        \(
            (\ctx_\theta, \vec{r}_1)
            \gets
            \text{\SampleFixedWeightVect{$\ctx_\theta$, $\Ring_{\omega_r}$}}
        \)
        \;
        \(\vec{u} \gets \vec{r}_1 + \vec{h}\cdot\vec{r}_2\)
        \;
        \(
            \vec{c} 
            \gets 
            \text{\Encode{\vec{m}}}
            +
            \text{\Truncate{$\vec{s}\cdot\vec{r}_2 + \vec{e}$, $\ell$}}
        \)
        \;
        \Return{\((\vec{u},\vec{v})\)}
        \;
    }
\end{algorithm}
\end{minipage}\hfill
\begin{minipage}[t]{0.48\textwidth}
\begin{algorithm}[H]
    \caption{\hqc KEM Encapsulation}
    \label{alg:HQC-KEM.Encaps}
    \scriptsize
    \Function{\KEMEncaps{\ekKEM}}{
        \KwIn{
            Encapsulation key \ekKEM
        }
        \KwOut{32-byte shared secret key $K$, ciphertext \cKEM}
        \(\vec{m} \gets [0,256)^{k}\)
        \;
        \(\salt \gets [0,256)^{16}\)
        \tcp*{16-byte random salt}
        \((K,\theta) \gets G(H(\ekKEM\parallel\vec{m}\parallel\salt))\)
        \;
        \(\cPKE \gets \PKEEncrypt{\ekKEM, $\vec{m}$, $\theta$}\)
        \;
        \(\cKEM \gets (\cPKE, \salt)\)
        \;
        \Return{\((K, \cKEM)\)}
        \;
    }
\end{algorithm}
\end{minipage}

\noindent
\begin{minipage}[t]{0.48\textwidth}
\begin{algorithm}[H]
    \caption{\hqc PKE Decryption}
    \label{alg:HQC-PKE.Decrypt}
    \scriptsize
    \Function{\PKEDecrypt{\dkPKE, $(\vec{u},\vec{v})$}}{
        \KwIn{
            Decryption key \ekPKE, ciphertext \((\vec{u},\vec{v})\)
        }
        \KwOut{Message \vec{m}}
        \(\seedPKEdk \gets \dkPKE\)
        \;
        \(\ctxPKEdk \gets \text{\XOFInit{\seedPKEdk}}\)
        \;
        \(
            (\ctxPKEdk, \vec{y}) 
            \gets
            \text{\SampleFixedWeightVect{\ctxPKEdk, $\Ring_\omega$}}\label{line:leakage_kem}
        \)
        \;
        \(\vec{m} \gets \text{\Decode{$\vec{v} - \text{\Truncate{$\vec{u}\cdot\vec{y}$, $\ell$}}$}}\)
        \;
        \Return{\vec{m}}
    }
\end{algorithm}
\end{minipage}\hfill
\begin{minipage}[t]{0.48\textwidth}
\begin{algorithm}[H]
    \caption{\hqc KEM Decapsulation}
    \label{alg:HQC-KEM.Decaps}
    \scriptsize
    \Function{\KEMDecaps{\dkKEM, \cKEM}}{
        \KwIn{
            Decapsulation key \dkKEM, ciphertext \cKEM
        }
        \KwOut{Shared secret key $K'$}
        \((\ekKEM,\dkPKE,\sigma) \gets \dkKEM\)
        \;
        \((\cPKE, \salt) \gets \cKEM\)
        \;
        \(\vec{m}' \gets \text{\PKEDecrypt{\dkPKE, \cPKE}}\)
        \;
        \(
            (K',\theta')
            \gets
            G(H(\ekKEM\parallel\vec{m}'\parallel\salt))
        \)
        \;
        \(
            \cPKE'
            \gets
            \text{\PKEEncrypt{\ekKEM, $\vec{m}'$, $\theta'$}}
        \);
        \(\cKEM' \gets (\cPKE', \salt)\)
        \;
        \(\bar K \gets J(H(\ekKEM)\parallel\sigma\parallel\cKEM)\)
        \;
        \If{\(\vec{m}' = \bot\) or \(\cKEM' \not= \cKEM\)}{
            \(K' \gets \bar K\)
        }
        \Return{\(K'\)}
        \;
    }
\end{algorithm}
\end{minipage}

Table~\ref{tab:parameters} lists the parameters of each \hqc instance.
The values $n_1$ and $n_2$ are related to the code $\mathcal{C}$;
the vector length parameter $n$ is the smallest primitive prime greater
than $n_1n_2$.
The parameter $\omega$ specifies the Hamming weight of the secret
vectors: in \hqc-{1}, for example, $n = 17\,669$ and $\omega = 66$. 

\begin{table}[t]
    \centering
    \caption{Parameter sets for \hqc~\cite{gaborit2025hamming}.
             Here $n_1$ and $n_2$ are the external and internal code lengths;
             $n$ is the ambient vector length
             (note that $n$ is the smallest primitive prime $> n_1 n_2$);
             $k$ is the code dimension (bits);
             $\omega$ is the Hamming weight of secret vectors $(x,y)$;
             and
             $\omega_r = \omega_e$ are the Hamming weights
             of the ephemeral vectors $(r_1,r_2,e)$.}
    \label{tab:parameters}
    \begin{tabular}{lcccccccc}
        \toprule
        Instance & Security & $n_1$ & $n_2$ & $n$ & $k$ & $\omega$
                 & $\omega_r = \omega_e$ & DFR \\
        \midrule
        \hqc-1 & NIST-1 & $46$ & $384$ & $17\,669$ & $128$ & $66$
               & $75$  & $< 2^{-128}$ \\
        \hqc-3 & NIST-3 & $56$ & $640$ & $35\,851$ & $192$ & $100$
               & $114$ & $< 2^{-192}$ \\
        \hqc-5 & NIST-5 & $90$ & $640$ & $57\,637$ & $256$ & $131$
               & $149$ & $< 2^{-256}$ \\
        \bottomrule
    \end{tabular}
\end{table}
\paragraph{Fixed-weight sampling.}
During decapsulation, the secret key $\seed$ is expanded using a pseudorandom
generator to reproduce the vectors $x$ and $y$ via a deterministic procedure
\[
    y \leftarrow \SampleFixedWeightVect(\cdot).
\]
The vector $y$ is a long binary vector of length $n$ and very small Hamming
weight $\omega \ll n$.
This vector is then stored in memory and subsequently used in arithmetic
operations.

\subsection{Power side-channel analysis}
\label{subsec:sca-background}

Side-channel analysis (SCA) exploits physical quantities emitted by a device---power consumption, electromagnetic radiation, timing, or
acoustic emissions---rather than weaknesses in the underlying algorithm.
In a \emph{power} side-channel attack, the adversary measures the current
drawn by the target at each clock cycle.
The resulting waveform, called a \emph{power trace}, leaks information about
the data being processed because the dynamic power dissipated by CMOS logic
is proportional to the number of gate transitions, which depends on the data
handled by the processor at each step~\cite{KocherJJ99}.


In SPA, a single trace (or very few traces) suffices to distinguish between
execution paths or operand values because the leakage is apparent from direct
inspection~\cite{KocherJJ99}.
SPA is effective when different code paths produce clearly different power
profiles — the canonical example being the square-and-multiply exponentiation
in RSA, where multiplications and squarings yield visually distinct patterns.


\paragraph{Leakage models.}
The relationship between an intermediate value $v$ and the measured power
consumption $L$ is described by a \emph{leakage model}.
Three leakage models are relevant to our analysis.

First,
the Hamming weight (HW) model assumes that power consumption is
proportional to the number of bits set to one in the processed value:
\begin{equation}
\label{eq:hw-model}
    L = a \cdot \hw(v) + b + \varepsilon,
    \qquad
    \varepsilon \sim \mathcal{N}(0,\sigma^2),
\end{equation}
where $a$ and $b$ are device-dependent constants, and $\varepsilon$ represents
additive Gaussian measurement noise with mean $0$ and variance $\sigma^2$.
The HW model accurately captures the leakage behavior of many 8- and 32-bit
microcontrollers, where the instantaneous power consumption is closely related
to the number of active data lines~\cite{MangardOP07}.

For pipelined processors and hardware in
which power consumption is dominated by register or bus transitions,
a more accurate model assumes
that power consumption is proportional to the number of bits that change between
successive values:
\begin{equation}
\label{eq:hd-model}
    L = a \cdot \hw(v \oplus v_{\mathrm{prev}}) + b + \varepsilon
\end{equation}
where $v_{\mathrm{prev}}$ is the previous value of $v$ in the bus.

\paragraph{Zero-value model.}
Our attack relies on a special case of the Hamming
weight model, which we refer to as the \emph{zero-value leakage model}. We
assume that the power trace produced when processing an all-zero operand is
statistically distinguishable from the trace corresponding to any nonzero
operand, without requiring the exact Hamming weight of the operand to be
observable. Defining the indicator variable
\[
    z = \mathbf{1}[\hw(v) > 0],
\]
the leakage is modeled as
\begin{equation}
\label{eq:zero-model}
    L = \mu_0 + \Delta \cdot z + \varepsilon,
\end{equation}
where $\mu_0$ is the mean leakage for the zero operand and $\Delta$ is
the average increase in leakage for any nonzero operand.

\subsection{Leakage model for \hqc}
\label{subsec:leakage}

During decapsulation, $y$ is regenerated from the secret seed and consumed by
the polynomial-multiplication routine, which loads its machine words in a fixed
order (Section~\ref{sec:code_analysis}), yielding a sequence of power leakages
aligned with the individual word loads.

Fix a machine-word size $W\in\{32,64\}$, and let $m=\lceil n/W\rceil$.
Partition the secret vector into $W$-bit words:
$y=(y^{(0)},\dots,y^{(m-1)})$ with $y^{(j)}\in\mathbb{F}_2^W$, padding the last
word with zeros. As we only need to tell whether a word is zero, set $z_j=0$ if
$y^{(j)}=0$ and $z_j=1$ otherwise. Writing $L_{t,j}$ for the leakage of the
$j$-th word in trace $t$, the zero-value model gives
\begin{equation}
\label{eq:leakage-model-main}
    L_{t,j}=\mu_0+\Delta\,z_j+\eta_{t,j},
    \qquad \eta_{t,j}\sim\mathcal{N}(0,\sigma^2),
\end{equation}
with $\mu_0$ the mean leakage of a zero word and $\Delta$ the average increase
for a nonzero word. Because the private key is reused, $y$ is processed
identically on every call, so an adversary can average $T$ aligned traces,
$\bar{L}_j=\frac1T\sum_t L_{t,j}$, with $\mathbb{E}[\bar{L}_j]=\mu_0+\Delta z_j$
and $\operatorname{Var}(\bar{L}_j)=\sigma^2/T$: more traces sharpen the
zero/nonzero decision.
A classifier on $\bar{L}_j$ returns $\hat{z}_j\in\{0,1\}$.
The set of indices 
$J_0=\{\,0\le j<m:\hat{z}_j=0\,\}$
of words judged zero
determines the known-zero coordinate set
\[
    Z:=\bigcup_{j\in J_0}\bigl\{\,jW,\dots,\min\!\bigl((j{+}1)W-1,\,n-1\bigr)\bigr\},
\]
with complement $S:=\{0,\dots,n-1\}\setminus Z$ and $n'=|S|$ the remaining
unknown coordinates.

\subsubsection{Decoding with leakage hints}
\label{subsec:decoding-hints}

The known-zero set $Z$ restricts the secret to its unknown coordinates
$y' := (y_i : i \in S) \in \F_2^{n'}$. Multiplication by $h$ is the circulant
matrix $\Rot(h)\in\F_2^{n\times n}$, so
relation~\eqref{eq:original-relation} reads
$s^\top = x^\top + \Rot(h)\,y^\top$. Since $y_i=0$ for every $i\in Z$, only the
columns of $\Rot(h)$ indexed by $S$ contribute; writing
$\Rot(h)_S\in\F_2^{n\times n'}$ for that submatrix,
\begin{equation}
\label{eq:reduced-relation}
    s^\top = x^\top + \Rot(h)_S\,(y')^\top .
\end{equation}
The number of equations is unchanged, but the number of unknowns drops from $2n$ to $n+n'$:
the leakage modifies neither the code dimension nor the number of observations,
removing only the coordinates known to be zero. This yields a \emph{decoding
problem with hints},
\[
    H'(x\|y')^\top = s^\top,
    \qquad
    H' = \bigl[\, I_n \;\; \Rot(h)_S \,\bigr] \in \F_2^{n\times(n+n')},
\]
attackable by information-set decoding adapted to the known-zero positions,
with the smaller unknown $y'$ in place of $y$.

\begin{remark}
The reduction depends directly on the leakage quality: when a large fraction of
the coordinates of $y$ are identified as zero, $n'\ll n$ and the decoding
instance becomes substantially smaller.
\end{remark}

\paragraph{Effect of the word size.}
Since $\hw(y)=\omega\ll n$, most machine words are all-zero, and a smaller $W$
raises the probability that an entire word is zero. Reducing $W$ thus enlarges
$Z$, shrinks $n'$, and gives stronger hints and smaller decoding instances.

\section{Searching for leakage}
\label{sec:code_analysis}
To assess and exploit the leakage, we must locate
load and store operations on \ssvs in the reference
C code of \hqc\footnote{The code is available in \url{https://gitlab.com/pqc-hqc/hqc}, tag v5.0.0 (commit f46e5422).},
since there is no side-channel resistant implementation publicly available. 

As shown in Section~\ref{sec:prem}, there are two types of \ssvs in HQC: ones linked to a specific ciphertext
($\mathbf{r}_1, \mathbf{r}_2, \mathbf{e}$), i.e. regenerated at each
\KEMEncaps\footnote{Notation aligns with the 2025-08-22
specification of HQC:
\url{https://pqc-hqc.org/doc/hqc_specifications_2025_08_22.pdf}.}
routine, and ones linked to a key pair ($\mathbf{x}, \mathbf{y}$),
i.e. regenerated at each \KEMKeyGen routine.
In \refC, these \ssvs are manipulated
in the following functions:
\begin{itemize}
	\item vector addition in $\mathbb{F}_2^n$, corresponding to the C function \vadd ;
	\item vector multiplication in $\mathbb{F}_2^n$,
        corresponding to \vmul followed by \smul ;
	\item \texttt{SampleFixedWeightVect}, corresponding to 
        either \texttt{vect\_sample\_fixed\_weight1}
        or \texttt{vect\_sample\_fixed\_weight2},
        followed by \vsup.
\end{itemize}

Table~\ref{tbl:load_store_sparse} lists the functions loading and
storing these \ssv{}s.
There are multiple leakage
targets for a single \ssv in each HQC-KEM phase, for example:
in \KEMDecaps we have both a load and a store leakage of $\mathbf{y}$.

\begin{table}[h!]
    \centering
    \begin{tabular}{rr|l|l}
        \toprule
        & Vector & Loaded by C function & Stored by C function
        \\
        \midrule
        \multirow{2}{*}{\PKEKeyGen}
        & $\mathbf{x}$ & \vadd & \vsup 
        \\
        & $\mathbf{y}$ & \smul & \vsup 
        \\
        \midrule
        \multirow{3}{*}{\PKEEncrypt}
        & $\mathbf{r}_1$ & \vadd & \vsup
        \\
        & $\mathbf{r}_2$ & \smul (2$\times$) & \vsup
        \\
        & $\mathbf{e}$ & \vadd & \vsup
        \\
        \midrule
        \PKEDecrypt
        & $\mathbf{y}$ & \smul & \vsup 
        \\
        \bottomrule
    \end{tabular}
    \caption{Secret sparse vector loads and stores in the HQC reference implementation.}
    \label{tbl:load_store_sparse}
\end{table}

We present several leakage sources of secret data in the C implementation. However, our main target is 
\PKEDecrypt, i.e., the decapsulation process. During decapsulation, the implementation 
computes $\mathbf{u} \cdot \mathbf{y}$ using \smul, so the load leakage we exploit occurs during every 
decapsulation. Note that $\mathbf{y}$ is both stored, by \vsup when it is regenerated from the 
\seed, and loaded, by \smul, within a single decapsulation, providing two leakage options 
per call. Moreover, this represents a more realistic attack model.

\subsection{Analysis of target C functions}

We analysed \refC to identify memory instruction (load and store) leakages.
We are targeting Cortex-M4 microcontrollers,
so we compiled the code for an ARMv7E-M target
using GCC version 14.2.1, with the \texttt{Os} flag for size optimization
(a realistic embedded device scenario).
To qualitatively assess the load and store instructions leaking secret data,
we analyze the assembly code (obtained with \texttt{objdump})
corresponding to each of the 
C functions listed in Table~\ref{tbl:load_store_sparse}.

\paragraph{Analysis of \vadd.} \label{sec:vadd_ana}
Listing~\ref{lst:vadd} shows \vadd's source code from \refC.
We noticed that \ssvs
are always passed as the operand \texttt{v1}
in the calls listed in Table~\ref{tbl:load_store_sparse}.
The \emph{load} of the \ssv from memory
is highlighted in \highlight{yhl}{yellow}.
For comparison, we also highlight in
\highlight{rhl}{red} the result of the addition targeted by \cite{DBLP:journals/iacr/HuangWZY25}
in the case of $\mathbf{v} \oplus (\mathbf{uy})$.
(Their work applies a chosen-ciphertext DPA targeting
the output of the \vadd function processing two dense vectors.)

\begin{lstlisting}[style=smallC,caption={Constant-time vector addition in $\mathbb{F}_2^n$},label={lst:vadd}]
void vect_add(uint64_t *o, const uint64_t *v1, const uint64_t *v2, uint32_t size) {
    for (uint32_t i = 0; i < size; ++i) {
        (*@\highlight{rhl}{o[i]}@*) = (*@\highlight{yhl}{v1[i]}@*) ^ v2[i];
    }
}
\end{lstlisting}

Listing~\ref{lst:vadd_asm} gives an extract of the assembly code
generated by the compiler.
Notice the load double instruction
\texttt{ldrd}: it loads a 64-bit word into a low and a high 32-bit
register in 3 cycles.\footnote{According to the ARM documentation: \url{https://developer.arm.com/documentation/ddi0439/b/Programmers-Model/Instruction-set-summary/Cortex-M4-instructions}}
This gives us our first source of leakage.
With the \texttt{O3} optimization, the compiler
uses the same \texttt{ldrd} instruction for fetching \texttt{v1[i]} from memory, hence \texttt{Os} and
\texttt{O3} provide the same leakage source.

The low and high 32-bit values of \uint are loaded separately (as the
datapath on a Cortex-M4 is only 32-bits), but it is hard to distinguish
them without knowing the microarchitecture and having a very accurate
probe. Hence, we consider this operation as having a 64-bit word leakage
granularity: our distinguisher outputs a 64-bit value that is either zero or nonzero.

\begin{lstlisting}[style=smallARM,caption={Extract of the assembly of
\vadd, GCC 14.2.1 \texttt{-Os}},label={lst:vadd_asm}]
(*@\highlight{yhl}{ldrd \, r10, r11, [r1], \#8}@*) ; load low and high 32-bit values of v1[i] into r10 and r11
ldrd   r8, r9, [r2, #8]!
eor.w  r4, r10, r8
eor.w  r5, r11, r9
(*@\highlight{rhl}{strd \, r4, r5, [r0, \#8]!}@*)  ; store low and high 32-bit values of o[i]
\end{lstlisting}

\paragraph{Analysis of \smul.} \label{sec:smul_ana}
Vector multiplication in $\mathbb{F}_2^n$ is implemented with
Karatsuba's algorithm:
\vmul calls the recursive function \kmul, which calls \smul at each tail call.
As before,
\ssvs are always passed as the first operand to \vmul, \kmul, and \smul.
However, only slices of \ssvs are passed at each tail call to \smul.
We can map each \smul call to the indices of the \ssv 
because calls are deterministic and not randomized.

Listing~\ref{lst:smul} gives \smul's source code.
The load of the \texttt{a} operand,
corresponding to slices of the \ssv,
is highlighted in \highlight{yhl}{yellow}.
The mask computation and operations highlighted in \highlight{rhl}{red}
constitute an additional attack surface: a 64-bit mask set to
\texttt{0xFF...FF} indicates that \texttt{bit}-th bit of the \texttt{i}-th processed 64-bit word
of the current \ssv slice \texttt{a} is set to one. As this is rare, because of
the sparsity, one can implement an SPA and recover the \ssvs
\texttt{y} and \texttt{r2}, according to Table~\ref{tbl:load_store_sparse}.
This attack was first presented by~\cite{velek2026simplepoweranalysispolynomial},
on a slightly different implementation: the Additional Implementation of HQC submitted to the 4th round of the NIST PQC competition.
We successfully ported their SPA to \refC,
but this work focuses only on the load of \texttt{a[i]}.

\begin{lstlisting}[style=smallC,caption={Constant-time schoolbook multiplication in $\mathbb{F}_2^n$.},label={lst:smul}]
static void schoolbook_mul(uint64_t *r, const uint64_t *a, const uint64_t *b, size_t n) {
    memset(r, 0, 2 * n * sizeof(uint64_t));
    for (size_t i = 0; i < n; i++) {
        (*@\highlight{yhl}{uint64\_t ai = a[i];}@*)
        for (int bit = 0; bit < 64; bit++) {
            (*@\highlight{rhl}{uint64\_t mask = -((ai >> bit) \& 1ULL);}@*)
            size_t base = i;
            int sh = bit;
            int inv = 64 - sh;
            if (sh == 0) {
                for (size_t j = 0; j < n; j++) {
                    (*@\highlight{rhl}{r[base + j] \textasciicircum{}= b[j] \& mask;}@*)
                }
            } else {
                for (size_t j = 0; j < n; j++) {
                    (*@\highlight{rhl}{r[base + j] \textasciicircum{}= (b[j] << sh) \& mask;}@*)
                    (*@\highlight{rhl}{r[base + j + 1] \textasciicircum{}= (b[j] >> inv) \& mask;}@*)
                }
            }
        }
    }
}
\end{lstlisting}

Listing~\ref{lst:smul_asm} provides an extract of the assembly code for \smul.
The instructions corresponding to Line~4 of
Listing~\ref{lst:smul}---which manipulates secret data---are highlighted in \highlight{yhl}{yellow}.
Observe that the compiler preferred split loads (Lines 2 and 5) for the low and high 32-bit values
of \texttt{a[i]} over an \texttt{ldrd} instruction. This is because it has to account
for loop operations, like counter and address updates. Because of the split at the instruction
level, this target will have a 32-bit word leakage granularity. Moreover, because of register
spilling, there is a store (Line 4) of the low 32-bit value of \texttt{a[i]};
hence,
we foresee higher leakage of the low part than of the high part.
That is: a nonzero word having a \texttt{1} in the low 32-bits
(i.e. positions~1 to~32)
will leak more because the load is followed by a store,
while a word with a \texttt{1} in the high 32-bits,
(positions~33 to~64) will leak less because there is only a load.

\begin{lstlisting}[style=smallARM,caption={Extract of the assembly of
\smul, GCC 14.2.1 \texttt{-Os}},label={lst:smul_asm}]
ldr     r2, [sp, #36]  ; load the pointer address of a[i] from memory
(*@\highlight{yhl}{ldr.w \, r3, [r2, \#8]!}@*) ; load the low 32-bit values of a[i] into r3, increment i
str     r2, [sp, #36]  ; store the updated pointer address value for next iteration
(*@\highlight{yhl}{str \;\;\;\; r3, [sp, \#44]}@*) ; store the low 32-bit values of a[i] back (register spilling)
(*@\highlight{yhl}{ldr.w \, lr, [r2, \#4]}@*)  ; load the high 32-bit values of a[i] into lr (r14)
\end{lstlisting}

Listing~\ref{lst:smul_asmo3} gives the assembly output
when \smul is compiled with \texttt{O3} optimization.
Interestingly, 
we get a new store of the high 32-bit values because of register spilling, 
increasing the surface of the leakage. \emph{This is a concrete example where speed optimizations can increase
side-channel leakage of manipulated secrets}.
Indeed, the compiler optimization almost doubles the leakage of the secret value with the additional \texttt{str}.

\begin{lstlisting}[style=smallARM,caption={Extract of the assembly of
\smul, GCC 14.2.1 \texttt{-O3}},label={lst:smul_asmo3}]
ldr     r2, [sp, #56]  ; load the pointer address of a[i] from memory
(*@\highlight{yhl}{ldr.w \; r1, [r2, \#8]!}@*) ; load the low 32-bit values of a[i] into r3, increment i
str     r2, [sp, #56]  ; store the updated pointer address value for next iteration
(*@\highlight{yhl}{ldr \;\;\;\;\, r2, [r2, \#4]}@*)  ; load the high 32-bit values of a[i] into lr (r14)
(*@\highlight{yhl}{str \;\;\;\;\, r1, [sp, \#44]}@*) ; store the low 32-bit values of a[i] back (register spilling)
(*@\highlight{yhl}{str \;\;\;\;\, r2, [sp, \#48]}@*) ; store the low 32-bit values of a[i] back (register spilling)
\end{lstlisting}

\paragraph{Analysis of \vsup.}
Listing~\ref{lst:vsup} shows the C source code for \vsup,
which turns a support (a list of bit
positions) into a sparse vector,
with each 64-bit word updated highlighted in \highlight{yhl}{yellow}.
In \refC, \vsup is
always called with a zero vector \texttt{v}, and \texttt{val} is initialized to
0 (Line 11), so at the end of the loop \texttt{val} holds the \texttt{i}-th
element of the \ssv and the logical ``or'' is effectively an assignment storing
it to memory---our target. The mask computation highlighted in
\highlight{rhl}{red} provides a strong zero/nonzero indicator:
an inner-loop mask of \texttt{0xFF...FF} marks the \texttt{i}-th
element as nonzero.
An SPA analogous to~\cite{velek2026simplepoweranalysispolynomial},
which we validated on \smul, is therefore possible here too---but it would
recover only zero 64-bit words, with no simple way to locate the set bits, and
is otherwise identical to~\cite{velek2026simplepoweranalysispolynomial}, so we
dismiss this target.

\begin{lstlisting}[style=smallC,caption={Constant-time support-to-vector conversion.},label={lst:vsup}]
void vect_write_support_to_vector(uint64_t *v, uint32_t *support, uint16_t weight) {
    uint32_t index_tab[PARAM_OMEGA_R] = {0};
    uint64_t bit_tab[PARAM_OMEGA_R]   = {0};
    for (size_t i = 0; i < weight; i++) {
        index_tab[i] = support[i] >> 6;
        int32_t pos  = support[i] & 0x3f;
        bit_tab[i]   = ((uint64_t)1) << pos;
    }
    uint64_t val = 0;
    for (uint32_t i = 0; i < VEC_N_SIZE_64; i++) {
        val = 0;
        for (uint32_t j = 0; j < weight; j++) {
            (*@\highlight{rhl}{uint32\_t tmp = i - index\_tab[j];}@*)
            (*@\highlight{rhl}{int val1 = 1 \textasciicircum{} ((tmp \textbar{} -tmp) >> 31);}@*)
            (*@\highlight{rhl}{uint64\_t mask = -val1;}@*)
            (*@\highlight{rhl}{val |= (bit\_tab[j] \& mask);}@*)

        }
        (*@\highlight{yhl}{v[i] |= val;}@*)
    }
}
\end{lstlisting}

Listing~\ref{lst:vsup_asm} gives the assembly for \texttt{v[i] |= val}
(Line 19 of Listing~\ref{lst:vsup})
with the secret-storing instruction highlighted in \highlight{yhl}{yellow}.
As for \vadd,
storing a 64-bit value on a 32-bit architecture compiles to a store double
\texttt{strd} (identical under \texttt{O3}), so, as in
Section~\ref{sec:vadd_ana}, the leakage granularity is 64-bit.

\begin{lstlisting}[style=smallARM,caption={Extract of assembly of \vsup
from GCC 14.2.1 \texttt{-Os}},label={lst:vsup_asm}]
ldrd   r2, r3, [r0, #8]! ; load low and high 32-bit values of v[i] (which are 0)
orr.w  r9, r2, r6        ; logical OR between low 32-bit values of v[i] and val
orr.w  r10, r3, r7       ; logical OR between high 32-bit values of v[i] and val
(*@\highlight{yhl}{strd \; r9, r10, [r0]}@*)    ; store the updated low and high 32-bit values of v[i]
\end{lstlisting}

\subsection{Choice of the target C function}

As we mentioned above, \KEMDecaps is the most realistic attack
target.
Our analysis above shows that conveniently, the C \smul
function used by \KEMDecaps is also the target most vulnerable to
our methods.

Table~\ref{tbl:cfunc_asm} lists the assembly instructions that we can
target in each of the C functions analyzed above with \texttt{Os}
optimization,
each with their leakage granularity.
We chose to attack \smul because of the smaller
leakage granularity: that is, we can distinguish zero versus nonzero words of size 32-bits
instead of the 64-bit words in the other two functions.
Moreover, since \smul has
both load and store operations, this choice enables
a leakage characterization of loads versus stores.
Another interesting aspect
in the context of \texttt{Os} optimization is the asymmetric
leakage between the low 32-bit values of the 64-bit \ssv word, which are loaded and stored,
versus the high 32-bit values, which are only loaded.
Future work can quantify the leakage of operations \texttt{ldrd} and \texttt{strd} coming from the remaining two functions.

\begin{table}[h!]
    \centering
    \begin{tabular}{ c|c|c }
        \toprule
        C function & Target assembly instructions & Leakage granularity
        \\
        \midrule
        \vadd & \texttt{ldrd} & 64-bit word
        \\
        \smul & \texttt{ldr.w, str, ldr.w} & 32-bit word
        \\
        \vsup & \texttt{strd} & 64-bit word
        \\
        \bottomrule
    \end{tabular}
    \caption{Target assembly instructions in each target C function}
    \label{tbl:cfunc_asm}
\end{table}

\section{A theoretical attack}
\label{sec:theory}
Recall from Section~\ref{sec:prem} that during KEM decapsulation the
secret vector~$y$ is deterministically regenerated from the secret \seed,
stored in memory, and consumed by the schoolbook multiplication routine,
which iterates over the 64-bit words $a[i]=y^{(j)}$ of~$y$ and loads each
word before the inner mask--accumulate step
(Section~\ref{sec:code_analysis}, Listing~\ref{lst:smul}).
Our goal is to recover~$y$ (and hence the full secret key) from the power
leakage of these word loads: the distinguisher below classifies each
$y^{(i)}$ as zero or nonzero, producing
known-zero-position hints for the decoding-with-hints framework.

\subsection{Leakage model}
\label{subsec:leakage-model}

The channel we exploit is the load of the secret word $a[i]$ in
\texttt{schoolbook\_mul}, not the per-bit mask computation. On the target
ARMv7E-M core (Cortex-M4) a 64-bit word is transferred as two 32-bit
register operations, and under \texttt{-Os} the low half is additionally
spilled to the stack, so the low half incurs a load \emph{and} a store while
the high half incurs only a load. We therefore model each half separately.

Write $y^{(j)}\in\FF_2^{64}$ for the $j$-th word of~$y$ and let
$y^{(j)}_{\mathrm{lo}},y^{(j)}_{\mathrm{hi}}\in\FF_2^{32}$ be its low and
high 32-bit halves. Refining the word-level
model~\eqref{eq:leakage-model-main} to the two halves, we introduce, for
each half $h\in\{\mathrm{lo},\mathrm{hi}\}$, the indicator
\begin{equation}
    z_j^{(h)} =
    \begin{cases}
        0, & y^{(j)}_{h}=0,\\
        1, & \text{otherwise,}
    \end{cases}
    \label{eq:halfword-indicator}
\end{equation}
and model the leakage sampled while word~$j$ is processed in trace~$t$ as
\begin{equation}
    L_{t,j}^{(h)}
    = \mu_0^{(h)} + \Delta^{(h)}\, z_j^{(h)} + \eta_{t,j}^{(h)},
    \qquad \eta_{t,j}^{(h)}\sim\mathcal{N}(0,\sigma^2),
    \label{eq:halfword-leakage}
\end{equation}
with $\mu_0^{(h)}$ the mean leakage of a zero half and $\Delta^{(h)}$ the
average increase for a nonzero one. Since the low half is both loaded and
stored while the high half is only loaded,
and also because stores leak more than loads,
$\Delta^{(\mathrm{lo})} > \Delta^{(\mathrm{hi})}$: the low half is the
stronger channel. This asymmetry, predicted by the compiled
code, is confirmed experimentally in Section~\ref{sec:exp}: the low
half can be distinguished with an order of magnitude fewer traces.
Under \texttt{-O3} the high half is spilled as well 
(Listing~\ref{lst:smul_asmo3}),
raising $\Delta^{(\mathrm{hi})}$ toward $\Delta^{(\mathrm{lo})}$.

\paragraph{Trace averaging.}
Since the same private key $y$ is reused,
an adversary acquires $T$ aligned traces and averages
$\bar{L}_j^{(h)}=\frac1T\sum_t L_{t,j}^{(h)}$, with
$\mathbb{E}[\bar{L}_j^{(h)}]=\mu_0^{(h)}+\Delta^{(h)}z_j^{(h)}$ and
$\mathrm{Var}(\bar{L}_j^{(h)})=\sigma^2/T$. The variance decreases at
rate $1/T$, so both halves are eventually classifiable; the weaker high
half simply requires proportionally more traces.

\paragraph{Word-level indicator.}
A 64-bit word is zero if and only if both halves are zero, so
$z_j = z_j^{(\mathrm{lo})}\lor z_j^{(\mathrm{hi})}$. Classifying at 32-bit
granularity thus subsumes the 64-bit classification while producing strictly
more zero-coordinate information.

\paragraph{Sparse occupancy.}
Because $y$ is extremely sparse, most machine words are identically zero.
Partitioning $y\in\{0,1\}^n$ of weight~$\omega$ into $m=\lceil n/W\rceil$
words of $W$ bits, a standard sparse-occupancy approximation gives
\begin{equation}
    \Pr[\text{word}=0]
    \approx \left(1-\frac{W}{n}\right)^{\omega}
    \approx \exp\!\left(-\frac{\omega W}{n}\right).
    \label{eq:zero-prob}
\end{equation}
For HQC-1 ($n=17\,669$, $\omega=66$) this is $\approx 0.787$ for $W=64$
and $\approx 0.887$ for $W=32$:
the half-word channel thus resolves positions more finely
and eliminates a larger fraction of coordinates.

\subsection{Zero-word distinguisher}
\label{subsec:zero-word-distinguisher}

For each (half-)word, the attacker compares its averaged leakage against a
template built from processing a zero word, using a statistical test: a
significant deviation classifies the (half-)word as nonzero, otherwise it is
declared zero, yielding $\hat{z}_j^{(h)}\in\{0,1\}$ and hence~$\hat{z}_j$.
The asymmetry $\Delta^{(\mathrm{lo})}>\Delta^{(\mathrm{hi})}$ gives two
regimes:
\begin{itemize}
    \item \textbf{Word granularity ($W=64$).}
        Classifying whole words only
        requires detecting a nonzero low \emph{or} high half; the strong low-half
        channel dominates, so 64-bit words are separated with the fewest traces.
    \item \textbf{Half-word granularity ($W=32$).} 
        Distinguishing the halves separately requires the weaker high half,
        costing an order of magnitude more traces
        but increasing the expected known zero-coordinate fraction $\rho$ from
        $\approx0.787$ to $\approx0.887$.
\end{itemize}
\begin{remark}
    The distinguisher resolves \emph{zero versus nonzero} at 32-bit granularity;
    it does not reveal the positions of the set bits inside a nonzero half-word.
    Locating the support within nonzero words is left to the
    ``decoding-with-hints'' step.
\end{remark}

\subsection{Reduction to a shortened syndrome decoding instance}
\label{subsec:reduction}

Writing the public equation in vector form, let
$\Rot(h)\in\FF_2^{n\times n}$ be the circulant matrix corresponding to~$h$, so
\begin{equation}
    s^\top = x^\top + \Rot(h)\,y^\top.
    \label{eq:public}
\end{equation}
With $e=(x\,\|\,y)\in\FF_2^{2n}$ and $H=[I_n\mid\Rot(h)]\in\FF_2^{n\times 2n}$,
this is the structured syndrome decoding instance
\begin{equation}
    H\,e^\top = s^\top,
    \qquad \hw(x)=\hw(y)=\omega.
    \label{eq:sdi}
\end{equation}

Let $J_0$ be the word indices classified as zero, inducing the
zero-coordinate set
\begin{equation}
    Z \supseteq \bigcup_{j\in J_0}\{jW,\dots,\min((j+1)W-1,\,n-1)\},
    \label{eq:Z}
\end{equation}
where $W\in\{32,64\}$ is the classification granularity. Let
$S=\{0,\dots,n-1\}\setminus Z$, $n'=|S|$, and $y'\in\FF_2^{n'}$ the
restriction of~$y$ to~$S$. Substituting $y_i=0$ for all $i\in Z$
into~\eqref{eq:public} and letting $\Rot(h)_S$ keep only the columns indexed
by~$S$,
\begin{equation}
    s^\top = x^\top + \Rot(h)_S\,(y')^\top.
    \label{eq:reduced-eq}
\end{equation}
Defining the shortened parity-check matrix
\begin{equation}
    H' = \bigl[I_n\;\;\Rot(h)_S\bigr]\in\FF_2^{n\times(n+n')},
    \label{eq:Hprime}
\end{equation}
this becomes the \emph{shortened} instance
\begin{equation}
    H'\,(x\,\|\,y')^\top = s^\top,
    \qquad \hw(x)=\hw(y')=\omega.
    \label{eq:short-sdi}
\end{equation}

The number of unknowns shrinks from $2n$ to $n+n'$.
Writing $n'\approx(1-\rho)n$ when a
fraction~$\rho$ of the coordinates of~$y$ is classified as zero, the
\emph{total} unknown length contracts by
\begin{equation}
    \frac{2n}{n+n'} \approx \frac{2}{2-\rho},
    \label{eq:reduction-factor}
\end{equation}
i.e. $\approx 1.80\times$ for HQC-1 at $W=32$ ($\rho\approx0.887$).
Equivalently, the $y$-block alone shrinks from~$n$ to $n'\approx0.113n$, a
factor $1/(1-\rho)\approx 8.9\times$ on the sparse unknown---the quantity
that drives the statistical-decoding gain of
Section~\ref{subsec:stat-decoding}. The constraint
$\omega(x)=\omega(y')=\omega$ is preserved, so every ISD technique exploiting
the two-block weight structure carries over.

\subsection{Robustness to classification errors}
\label{subsec:robustness}

The reduction assumes every coordinate placed in~$Z$ is truly zero. The two
classification errors are \emph{not} symmetric:
\begin{itemize}
    \item A \emph{false nonzero} (a zero word declared nonzero) merely fails
    to remove a coordinate: $n'$ is slightly larger and decoding marginally
    slower, but the true solution is untouched.
    \item A \emph{false zero} (a nonzero word declared zero) places a genuine
    support coordinate of~$y$ into~$Z$. Then $\hw(y')<\omega$ over~$S$, the
    true solution leaves the search space of~\eqref{eq:short-sdi}, and the
    decoder fails silently, since the weight check on~$x$ is never met.
\end{itemize}
Only false zeros are fatal, and they are one-sided, so the attacker biases the
distinguisher toward ``nonzero'': a (half-)word is declared zero only when its
averaged leakage is statistically \emph{indistinguishable} from the zero
template, rather than at the symmetric TVLA threshold. This suppresses false
zeros at the cost of harmless false nonzeros, i.e.\ a slightly larger~$n'$.

Quantitatively, the attack succeeds only if \emph{every} nonzero word is
caught. At $W=32$, HQC-1 has $m=\lceil 17\,669/32\rceil = 553$ half-words, of
which $m(1-\rho)\approx 62$ are nonzero in expectation. If $p$ is the per-word
false-zero probability, the expected number of lost support coordinates is
$\approx 62\,p$, so single-shot success requires $p\lesssim 1/62\approx1.6\times10^{-2}$,
comfortably $p\le10^{-3}$. Since $\mathrm{Var}(\bar L_j^{(h)})=\sigma^2/T$,
this is met by acquiring enough traces on the weaker high half; the reported
$T$ of Section~\ref{sec:exp} is chosen accordingly.

\begin{remark}
A residual handful of false zeros need not be fatal if the decoder is run with
a small weight slack: allowing $\hw(y')\in\{\omega-\delta,\dots,\omega\}$ over a
support enlarged by the suspected-zero coordinates re-admits the lost positions
at a modest increase in~$n'$, turning a hard failure into a graceful cost
increase whenever the high-half channel is trace-limited.
\end{remark}

\subsection{ISD-style key recovery after leakage}
\label{subsec:isd-after-leakage}

Starting from~\eqref{eq:short-sdi}, the block structure of $H'$ permits a
simple recovery. For any candidate $y'\in\FF_2^{n'}$ of weight~$\omega$, the
left block $I_n$ uniquely determines
\begin{equation}
    x = s + \Rot(h)_S\,y',
    \label{eq:x-from-yprime}
\end{equation}
so $y'$ is valid if and only if $\omega(x)=\omega$. Key recovery reduces to searching for
a weight-$\omega$ vector $y'$ over $n'$ positions subject to a single weight
check on~$x$, e.g. via meet-in-the-middle or block-wise
enumeration~\cite{DHorlemannPRSW21,DAchilleEK26}. The leakage has transformed
an instance of length~$2n$ into one of length $n+n'\approx 1.113n$ (HQC-1,
32-bit), making the combinatorial search substantially more feasible.

\begin{remark}
To make the effect concrete, we solved~\eqref{eq:short-sdi} with plain Prange
ISD on scaled quasi-cyclic instances ($n$ prime, $x,y$ of weight~$\omega$);
the expected iteration count $\binom{n+n'}{2\omega}\big/\binom{n}{2\omega}$
shrinks rapidly as leakage removes coordinates. On a toy instance ($n=211$,
$\omega=6$) revealing the zero words at $W=32$ gave a $17\times$ speed-up, and
$W=16$ a $35\times$ one, matching the combinatorial prediction. Extrapolating
to HQC-1 ($n=17\,669$, $\omega=66$), the expected Prange work falls from
$\approx 2^{134}$ (no leakage) to $\approx 2^{39}$ at $W=64$ and $\approx 2^{22}$
at $W=32$; plain Prange is used only to expose the scaling, and dedicated ISD
lowers these figures further.
\end{remark}

\subsection{Statistical decoding after leakage}
\label{subsec:stat-decoding}

The shortened instance also \emph{admits} statistical decoding as an
alternative to ISD, on the same instance~\eqref{eq:short-sdi}: rather than
enumerate, one recovers the support of~$y'$ from a bias in random linear
projections. Only the ISD route is implemented and benchmarked here
(Table~\ref{tab:complexity}); the sketch below shows the instance is amenable
to soft-decision techniques~\cite{DAchilleEK26}, but a complete decoder is
left to future work.

Since $h$ is uniform and $n$ prime, $h$ is invertible in $\Ring$ with
overwhelming probability~\cite{gaborit2025hamming}. Multiplying
\eqref{eq:reduced-eq} by $h^{-1}$ yields
\begin{equation}
    t^\top = (h^{-1}\cdot x)^\top + (y')^\top,
    \label{eq:stat-eq}
\end{equation}
where $t=h^{-1}\cdot s$ is publicly computable, $h^{-1}\cdot x$ is
random-looking noise, and $y'$ is the sparse signal. For any $\ell\in\FF_2^{n'}$
and index $i$, $\langle\ell,t\rangle=\langle\ell,y'\rangle+\langle\ell,h^{-1}x\rangle$;
the noise term has bias proportional to $\omega/n'$, and conditioning on
$\ell_i=1$ gives a detectable excess probability $(1-2\omega/n')/2$ for
$\langle\ell,t\rangle=y'_i$~\cite{DAchilleEK26}. Averaging over many random
$\ell$ with $\ell_i=1$ amplifies this into a test separating support
coordinates of~$y'$ from zero ones.

The leakage reduces the ambient dimension from~$n$ to~$n'$: for HQC-1 at $W=32$
the support density $\omega/n'$ rises by $1/(1-\rho)\approx 8.9$, concentrating
the sparse signal and shrinking the search space. With enough traces the noise
averages out, recovering the support of~$y'$; $x$ then follows uniquely
from~\eqref{eq:x-from-yprime}.

\begin{remark}
The advantage of the hints is a reduction of the ambient dimension, not an
amplification of a single projection's bias: the elementary per-coordinate bias
$(1-2\omega/n')$ in fact \emph{decreases} as $n'$ shrinks. Since the noise
$h^{-1}x$ is per-coordinate unbiased, a concrete statistical decoder must
exploit genuine low-weight parity checks, which we leave to future work.
\end{remark}


\section{Experimental Results}
\label{sec:exp}
Our target is an STM32F446RE, a platform not evaluated in~\cite{miracle}. The
assembly of \smul also differs from the kernels they construct: their closest
kernel, LD-ST, matches the assembly of \vsup, but in our case each operation acts
on two 32-bit registers holding one 64-bit C variable, whereas theirs manipulate a
single 32-bit register. We therefore cannot transfer their conclusions directly, though
they indicate that the leakage we target is plausible.

\subsection{Experimental Setup}

We target an STM32F446RE Nucleo board (Cortex-M4, clocked at 30\,MHz). The firmware
contains the targeted function \smul, compiled with the size optimization \texttt{Os}
using PlatformIO Core 6.1.18 (ARM GNU toolchain 14.2.1). We insert two software
triggers wrapping the assembly code of Listing~\ref{lst:smul_asm} to delimit the
region of interest without additional signal processing.

To obtain realistic operands, we compile \refC, tagged as release v5.0.0 (commit
f46e5422)\footnote{The latest version (commit 161cd4fd) does not change any of the
three functions analyzed in Section~\ref{sec:code_analysis}.},
adding \texttt{printf}s
to log the operands of the \smul calls handling the \ssvs \texttt{y} and
\texttt{r2} (Table~\ref{tbl:load_store_sparse}).
Running multiple HQC instances on an x86
machine yields a set of sample vectors, which we replay on the target while capturing
its electromagnetic (EM) emanations. We use a Langer RF-B 0,3-3 probe placed over a
region of the Cortex-M4 identified by prior EM cartography, two Langer PA~303
amplifiers (3\,GHz, 30\,dB), and a Teledyne LeCroy WP404HD oscilloscope sampling at
1\,GS/s.

\subsection{Leakage assessment}

A fine-grained assessment would distinguish the leakage of each individual bit, since
even within one 32-bit register bit~1 and bit~3 leak differently. Because we capture
EM emanations with a relatively large probe (3\,mm), we adopt the simplification that
all bits in the same 32-bit register share one leakage profile. The two 32-bit
registers are used in \emph{separate} instructions, so from the \uint perspective the
cumulative leakage of bits~1-32 (the low register) is separable from that of
bits~33-64 (the high register): the two contributions occur at distinct instants,
as each half is loaded or stored in a separate instruction. Moreover, as shown in
Section~\ref{sec:smul_ana}, the low half (a load \emph{and} a store) leaks with
significantly higher amplitude than the high half (a load only).
These two criteria---\emph{temporal occurrence} and \emph{amplitude}---let
the distinguisher attribute a
detected nonzero pattern to the low or the high half, effectively giving a 32-bit
leakage granularity.

Our goal is to separate a zero 64-bit word processed by \smul's main loop from the
low and high halves of a nonzero one. We therefore build three sparse vectors:
\begin{itemize}
    \item a zero vector;
    \item a vector with a single one at position~1 (processed in the low register);
    \item a vector with a single one at position~33 (processed in the high register).
\end{itemize}
We multiply each by a random ciphertext \texttt{u}, capture the EM leakage, and run
two Test Vector Leakage Assessments (TVLAs): the zero vector against the low-register
nonzero vector, and against the high-register nonzero vector. We expect the first to
reach significance with fewer traces, since there both \texttt{ldr.w} and \texttt{str}
leak, whereas only \texttt{ldr.w} leaks in the second (Section~\ref{sec:smul_ana}).

\begin{figure}[t]
    \centering
    \includegraphics[width=\linewidth]{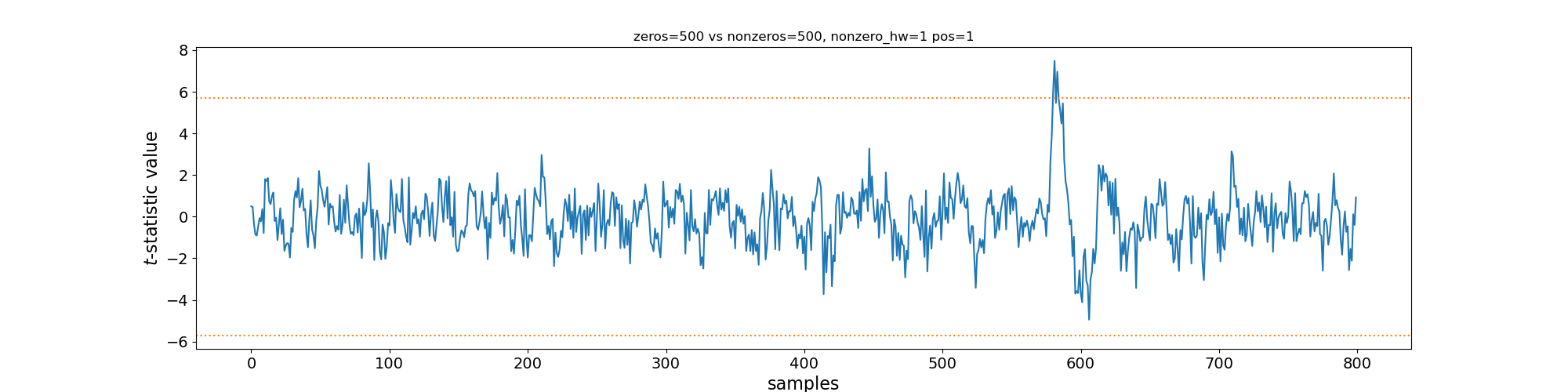}
    \includegraphics[width=\linewidth]{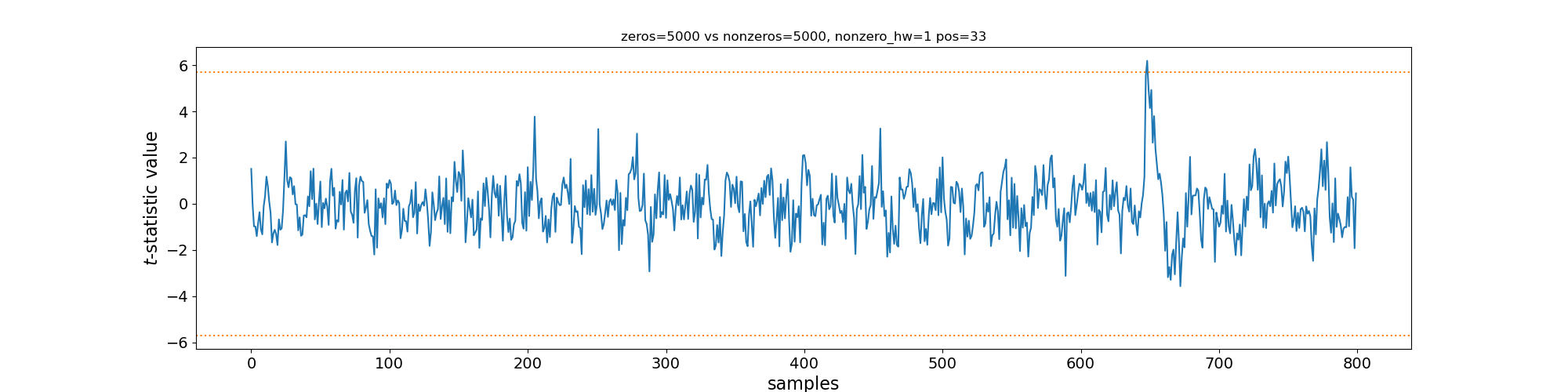}
    \caption{TVLA of zero vs.\ nonzero for bit position~1 with 500 traces per group
    (top) and bit position~33 with 5\,000 traces per group (bottom).
    The title of each graph indicates the number of traces for a zero \uint, the number
    of traces for a nonzero, its hw, and the position of ones (1 to 64) in it.
}
    \label{fig:tvla_smul}
\end{figure}

Figure~\ref{fig:tvla_smul} confirms this: the first TVLA crosses the threshold at
$\approx500$ traces and the second at $\approx5\,000$, so the low register leaks
markedly more than the high one, matching the \texttt{Os} assembly of
Listing~\ref{lst:smul_asm}. The TVLA threshold includes a correction over the $L$
samples of the segment following~\cite{DBLP:conf/asiacrypt/WhitnallO19}.

\begin{figure}[t]
    \centering
    \includegraphics[width=\linewidth]{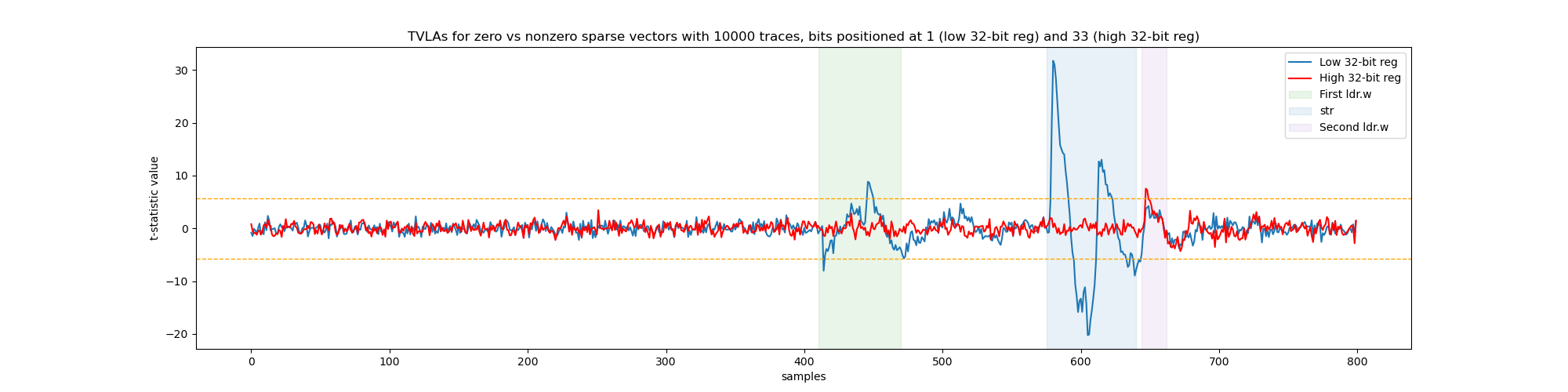}
    \caption{Superposed TVLAs over 10\,000 traces for zero vs.\ nonzero sparse vectors,
    with the nonzero part processed in the low register (blue) and in the high register
    (red). Highlighted sections map the leakage regions to the leaking instructions.}
    \label{fig:ldr_vs_str}
\end{figure}

At an equal trace count, the two TVLAs together characterise \texttt{ldr.w} versus
\texttt{str} for a HW-1 nonzero word relative to a zero one. Figure~\ref{fig:ldr_vs_str}
shows that on the STM32F446RE a memory write leaks more than a memory read; we map the
leakage regions to the responsible instructions by highlighting, confirming the
difference in both temporal occurrence and amplitude at equal traces.

\subsection{Practical attack}

Our attack is a \emph{profiled} template attack. 
To build zero templates, we run \smul on a zero sparse vector on the target device, then use the Welch $t$-test as the
distinguisher. Because the leakage is strong and easily averaged over the reference
implementation, plain zero templates suffice; Gaussian templates or more elaborate
methods are unnecessary. Such methods may reduce the trace count and improve leakage
exploitation on more noisy targets. We leave these left to future work as the 
distinguisher here already establishes our point: sparse vectors leak heavily
through memory operations.

For a target HQC-1 execution, we capture many traces and, at each loop iteration,
$t$-test the target \uint against the zero template of the corresponding iteration
(avoiding noise from the memory address value leakage): a positive result marks it
nonzero, otherwise zero. Since the \smul iterations map deterministically to \uint
positions in the \ssv (Section~\ref{sec:code_analysis}), all words are classified in
parallel, and the bottleneck is only the trace count needed for a positive $t$-test:
at worst $\approx5\,000$, for a nonzero confined to the high register.
This recovers the zeros of a \ssv at 32-bit granularity.

The trace count is the same
across all three HQC parameter sets, as they call the same \smul; run in parallel, the
attack time is also identical, the only differences being the per-trace acquisition
time (which grows with the \vmul length) and the number of parallel classifications
(which grows with the \smul iteration count).

\begin{figure}[t]
    \centering
    \includegraphics[width=\linewidth]{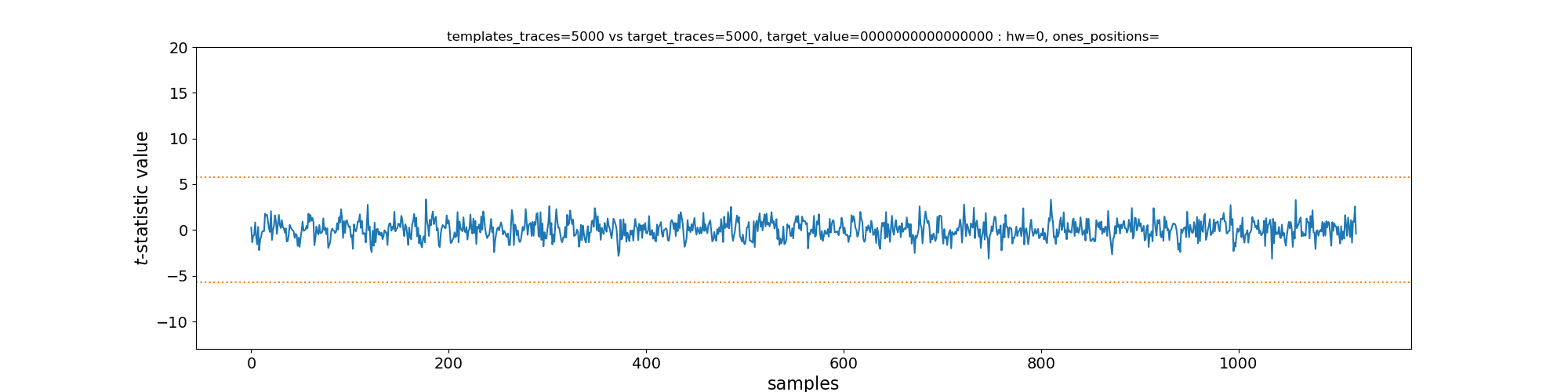}
    \includegraphics[width=\linewidth]{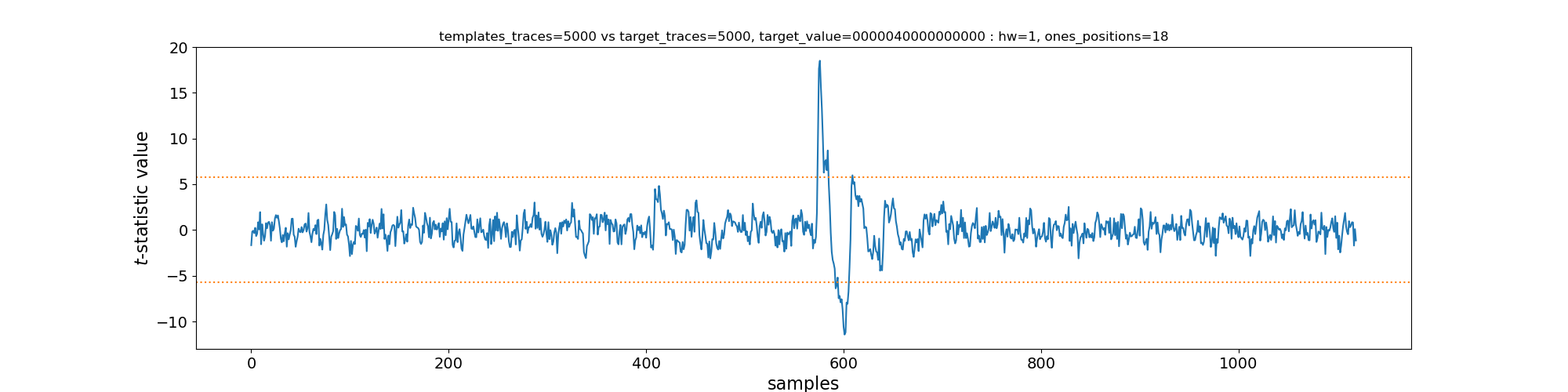}
    \caption{Attack on \texttt{y} being processed in \smul, iteration 2 (top) and iteration 3 (bottom).
    The title of each graph indicates the number of zero template traces, the number of target traces, the target
    value in little-endian hexadecimal format, its hamming weight, and the position of ones (1 to 64) in it, if any.
    }
    \label{fig:attack_smul}
\end{figure}

Figure~\ref{fig:attack_smul} shows \uint indices~2 and~3 (of $\{0,\dots,276\}$) for a
\texttt{y} from an HQC-1 instance: the word at index~2 is entirely zero,
while at index~3 the low half is nonzero and the high half is zero.
Over the whole vector we classify all
$m=553$ 32-bit half-words of \texttt{y}, of which \nnz{} are nonzero; we thus
classify $\approx\zerofrac\%$ of the coordinates to be zero, in line
with the expected $88.7\%$
of Section~\ref{subsec:leakage-model}. The nonzero half-words are then resolved by the
decoder of Section~\ref{subsec:complexity}. The threshold is the conservative,
false-zero-suppressing choice of Section~\ref{subsec:robustness}, since a single
misclassified nonzero word would drop a true support coordinate and break recovery.

\subsection{Remaining attack complexity}
\label{subsec:complexity}

The end-to-end cost of the attack decomposes into three parts:
\begin{enumerate}
    \item acquiring the EM traces,
    \item zero-word classification, and
    \item decoding the shortened instance~\eqref{eq:short-sdi}.
\end{enumerate}
We treat
each in turn, summarising the figures for each parameter set in
Table~\ref{tab:complexity}.

\paragraph{Trace acquisition.}
The distinguisher classifies every (half-)word of~$y$ in parallel from the
\emph{same} batch of aligned traces: a single \texttt{schoolbook\_mul}
execution exposes the loads of all $m=\lceil n/W\rceil$ words, so the query
cost is the number~$T$ of traces, not $T\cdot m$. As reported in
Section~\ref{sec:exp}, $T\approx 5\,000$ traces suffice for a reliable
Welch-$t$ decision in the worst case (a nonzero coefficient confined to the
weaker high half); the strong low half is already separated at
$\approx 500$ traces. Because the same \texttt{schoolbook\_mul} routine is
used at all three security levels, $T$ is essentially independent of the
parameter set: only the acquisition time \emph{per trace} grows with the
length of \texttt{vect\_mul}.

\paragraph{Classification.}
Each of the $m$ (half-)words requires one Welch-$t$ test between its $T$
target traces and the zero-template traces over the $L$ samples of the
aligned segment, i.e.\ $O(m\,T\,L)$ arithmetic operations in total. For
HQC-1 this is a few times $10^{9}$ operations --- negligible next to the
acquisition and decoding, and trivially parallel over the $m$ words.

\paragraph{Decoding.}
We solve the shortened instance~\eqref{eq:short-sdi} with the
block-structured information-set decoder implicit
in~\eqref{eq:x-from-yprime}. Each iteration draws a set $R$ of $n'$ of the
$n$ equations, hypothesises $x_R=0$, and solves the $n'\times n'$ linear
system
\begin{equation}
    (\Rot(h)_S)_R\,(y')^\top = s_R^\top ,
    \label{eq:block-solve}
\end{equation}
accepting the candidate if and only if $\hw(y')=\hw\!\bigl(s+\Rot(h)_S\,y'\bigr)=\omega$.
In other words, a draw succeeds exactly when the $\omega$ support positions of~$x$ all fall
outside~$R$ and $(\Rot(h)_S)_R$ is invertible. The information set~$R$ holds
exactly $n'=|S|$ of the $n$ equations, so its complement has size
$n-n'=|Z|$; a fixed coordinate of~$x$ therefore lies outside~$R$ with
probability $(n-n')/n=|Z|/n=\rho$, the same zero fraction measured by the
distinguisher on~$y$. The support of~$x$ avoids~$R$ with probability
\begin{equation}
    \Pr[\,\supp(x)\cap R=\emptyset\,]
    = \frac{\binom{n-\omega}{n'}}{\binom{n}{n'}}
    = \prod_{i=0}^{\omega-1}\frac{n-n'-i}{n-i}
    \approx \Bigl(1-\tfrac{n'}{n}\Bigr)^{\!\omega}
    = \rho^{\,\omega},
    \qquad n'=(1-\rho)\,n .
    \label{eq:p-avoid}
\end{equation}
Here $\rho=|Z|/n$ is a property of the leakage on~$y$, whereas the event
concerns the support of~$x$; the identity $\rho=(n-n')/n$ that links them
holds only because the decoder draws as many equations as there are
unknowns, $|R|=n'=|S|$. The single approximation is the standard
$\prod_{i}(n-n'-i)/(n-i)\approx(1-n'/n)^{\omega}$, valid for
$\omega\ll n$; the final equality $(1-n'/n)^{\omega}=\rho^{\omega}$ is exact.
The invertibility of $(\Rot(h)_S)_R$ contributes a further factor
$\approx P_{\mathrm{inv}}=\prod_{i\ge 1}(1-2^{-i})\approx 0.29$ (measured
$\approx 0.33$ on our instances). The expected number of
iterations and the total decoding work are therefore
\begin{equation}
    \mathbb{E}[\text{iters}] \approx \frac{\rho^{-\omega}}{P_{\mathrm{inv}}},
    \qquad
    T_{\mathrm{dec}} \approx \frac{\rho^{-\omega}}{P_{\mathrm{inv}}}\;\cdot\;(n')^{3}
    \ \ \text{bit operations},
    \label{eq:tdec}
\end{equation}
where $(n')^{3}$ is the bit cost of the one Gaussian elimination
in~\eqref{eq:block-solve}. We count \emph{bit} operations in
$T_{\mathrm{dec}}$; on a $w$-bit machine the elimination is run on packed
rows and costs $(n')^{3}/w$ \emph{word} operations, so a plain $64$-bit
implementation already saves a factor~$w=64$ over the bit count (for HQC-1
at $W=32$, $T_{\mathrm{dec}}=2^{46.1}$ bit operations $=2^{40.1}$ word
operations). Two features make this efficient: the identity block fixes~$x$
from~$y'$, so an iteration only requires the $\omega$ ones of~$x$ to avoid
$R$ (rather than \emph{all} $2\omega$ errors to avoid an information set),
which lowers the iteration count from $\rho^{-2\omega}$ to $\rho^{-\omega}$;
and the elimination in~\eqref{eq:block-solve} is $n'\times n'$ rather than
$n\times n$. For HQC-1 at $W=32$ this improves on the generic Prange bound
of Section~\ref{subsec:isd-after-leakage} from $2^{22}$ to $2^{13}$
iterations, each on a matrix a hundred times smaller.

\paragraph{Implementation and timing.}
Our reference
implementation stores the rows of $(\Rot(h)_S)_R$ as packed $64$-bit words
and performs one $n'\times n'$ elimination per iteration; on a single
core\footnote{AMD EPYC 7702 64-Core Processor (128 Threads) @ 3.35 GHz with 1TB RAM}
a \texttt{numpy}-based (i.e.\ unoptimised) solver takes $\approx 186$\,ms for
HQC-1 at $W=32$ ($n'=1997$), consistent with the $(n')^{3}/w\approx
1.2\times10^{8}$ word operations per iteration at the throughput of that
solver. Parallelism is process-level: $48$ workers run the randomised loop
of~\eqref{eq:block-solve} with independent seeds and the first to pass the
weight check halts the rest, so with $\mathbb{E}[\text{iters}]\approx2^{13.2}$
the expected wall-clock is $2^{13.2}\times186\,\text{ms}/48\approx40$\,s.
A dedicated $\FF_2$ linear-algebra kernel (e.g.\ the method-of-four-russians
solver of \textsc{m4ri}~\cite{M4RI}) would lower the per-iteration cost.

Table~\ref{tab:complexity} shows the complexity for a full
attack and the wall-clock figures. These wall-clock values are \emph{effective measurements},
not estimated by dividing $T_{\mathrm{dec}}$ by a clock rate. 
The decoding cost is dominated by~\eqref{eq:tdec} and depends
critically on the leakage granularity: the half-word hints obtained from
the \emph{combined store+load} channel ($W=32$, $\rho\approx0.89$) reduce the
decoding of HQC-1 to $2^{46}$ bit operations, which our parallel Python
implementation solves in about $40$~s on a $48$-core machine, whereas the load-only
whole-word hints ($W=64$, $\rho\approx0.79$) leave a $2^{60}$-operation
instance (several days on the same machine). The same $\approx5\,000$
traces feed both regimes, so the store leakage is precisely what turns an
impractical decoding into a routine one. At the higher security levels the
$W=32$ decoding stays feasible (minutes on a cluster), while the $W=64$
variant becomes impractical.

\begin{table}[t]
\centering
\caption{Attack complexity per parameter set and leakage granularity~$W$.
$\rho$: fraction of coordinates classified zero; $n'=|S|$: remaining
unknowns; $\mathbb{E}[\text{iters}]$ and $T_{\mathrm{dec}}$ (in \emph{bit}
operations) from~\eqref{eq:tdec}. The wall-clock column is
$\mathbb{E}[\text{iters}]\times$(measured per-iteration time)$/48$ for our
\texttt{numpy} solver, \emph{not} $T_{\mathrm{dec}}$ divided by a clock rate;
per-iteration times scale as $(n')^{3}$. The trace count $T\approx5\,000$ is
common to all rows.}
\label{tab:complexity}
\begin{tabular}{llcccccc}
\toprule
Instance & $W$ & $\rho$ & $n'$ & $\log_2\mathbb{E}[\text{iters}]$
         & $\log_2 T_{\mathrm{dec}}$ & wall-clock (48 cores) \\
\midrule
\multirow{2}{*}{HQC-1} & 64 & 0.787 & 3763 & 24.6 & 60.2 & $7.6$ d \\
                       & 32 & 0.887 & 1997 & 13.2 & 46.1 & $\approx 40$ s \\
\midrule
\multirow{2}{*}{HQC-3} & 64 & 0.837 & 5861 & 27.6 & 65.1 & $228$ d \\
                       & 32 & 0.915 & 3062 & 14.7 & 49.4 & $\approx 6$ min \\
\midrule
\multirow{2}{*}{HQC-5} & 64 & 0.865 & 7803 & 29.3 & 68.1 & $4.9$ y \\
                       & 32 & 0.930 & 4040 & 15.6 & 51.5 & $\approx 26$ min \\
\bottomrule
\end{tabular}
\end{table}

\section{Countermeasures}
\label{sec:counter}
\subsection{Additive masking}
\label{sec:masking}

One lightweight countermeasure is to never store~$y$ in its plain sparse form.
Instead, we can sample a fresh uniform-random $\mathbf{r} \in \mathbb{F}_2^n$ at each
decapsulation, storing the pair $(\mathbf{r},\, \mathbf{y}' := \mathbf{r}
\oplus \mathbf{y})$. The product $u \cdot y$ is then recovered as
\begin{equation}
  u \cdot y \;=\; u \cdot \mathbf{r} \;\oplus\; u \cdot \mathbf{y}',
\end{equation}
by computing two multiplications and XOR-ing the results.

As long as~$\mathbf{r}$ is freshly sampled per call, both operands are
computationally indistinguishable from uniform, hence non-sparse: each word is
zero with probability only $2^{-W}$, far below the $\approx 0.887$ rate
exploited by our distinguisher. The load attack therefore does not apply to
either multiplication, and the leakage is eliminated at its source. Note
that~$\mathbf{r}$ must be re-sampled every call: reusing it would expose a
fixed, non-sparse~$\mathbf{y}'$, letting an adversary average across
invocations, cancel~$\mathbf{r}$, and recover~$y$.

The cost is performance: decapsulation now requires two full-length polynomial
multiplications instead of one, roughly doubling the bottleneck operation, plus
a one-time cost to generate~$\mathbf{r}$ and to compute~$\mathbf{y}'$ by
flipping the $\omega$ support positions of~$y$ in~$\mathbf{r}$.
\subsubsection{Implementation and evaluation}
\label{sec:mask-eval}

We implemented the additive masking in the reference C implementation of HQC and 
measured its cost on all three parameter sets. The countermeasure is confined to the 
decapsulation product $u \cdot y$
computed in the \texttt{hqc\_pke\_decrypt} routine, and is selected at compile
time so that the unmodified reference remains available as a baseline. The mask
is drawn from the same DRBG the scheme already uses for the message and salt,
never from the secret key seed, and is freshly sampled on every decapsulation.

\paragraph{Two variants.}
We realise the mask in two ways.
\emph{(i) Per-call masking} follows Section~\ref{sec:masking} directly: at each
decapsulation a fresh $\mathbf{r} \in \mathbb{F}_2^n$ is sampled,
$\mathbf{y}' = \mathbf{y} \oplus \mathbf{r}$ is formed, and $u \cdot y = u \cdot
\mathbf{y}' \oplus u \cdot \mathbf{r}$ is evaluated with two multiplications and
an XOR. The mask is discarded afterwards, so the secret key is unchanged.
\emph{(ii) Stored-share masking with refresh} moves the split to key generation.
The secret key stores the two dense shares
$(\mathbf{y}' = \mathbf{y} \oplus \mathbf{r},\, \mathbf{r})$ in place of the
seed, so the sparse vector~$y$ is never reconstructed in the coefficient domain
during decapsulation. Each call first re-randomises the shares in place with a
fresh $\boldsymbol{\delta} \in \mathbb{F}_2^n$,
\begin{equation}
  \mathbf{y}' \leftarrow \mathbf{y}' \oplus \boldsymbol{\delta}, \qquad
  \mathbf{r}  \leftarrow \mathbf{r}  \oplus \boldsymbol{\delta},
\end{equation}
which preserves the invariant $\mathbf{y}' \oplus \mathbf{r} = y$ while changing
both stored words, so that no fixed secret-derived value is loaded twice across
calls, and then evaluates $u \cdot y = u \cdot \mathbf{y}' \oplus u \cdot
\mathbf{r}$ as above. This variant keeps the coefficient-domain vector~$y$
entirely off the repeatedly attacked path---it appears only transiently at key
generation---but it makes the decapsulation key \emph{stateful}, since the
refreshed shares are written back, and it enlarges the secret key, as two
$n$-bit vectors replace the seed. A stateful key must be held in
rollback-resistant storage: resetting it to a previous state would reinstate a
repeated mask and re-enable trace averaging. Both variants are first-order: they
remove the sparse single-share leakage the attack exploits, but not a
hypothetical second-order combination of the two shares. We verified that both
variants decapsulate correctly and return exactly the same shared secret as the
unmodified reference across all parameter sets.

\paragraph{Setup.}
We benchmark the reference C implementation compiled with GCC $16.1.1$ and with 
\texttt{-O3} on a 13th~Gen Intel Core i7-1365U. For each parameter set and mode we 
report the median cycle count, obtained with \texttt{rdtscp} over $2000$, $500$, and $200$
iterations for HQC-1, HQC-3, and HQC-5, respectively, at two granularities: the
isolated decryption routine \texttt{hqc\_pke\_decrypt} (the operation the
countermeasure protects) and the full decapsulation \texttt{crypto\_kem\_dec}.
As the reference multiplication is a schoolbook/Karatsuba routine rather than an
optimised kernel, the absolute counts are large; the \emph{ratios} to the
unmasked baseline are the implementation-independent quantities of interest.

\paragraph{Results.}
Table~\ref{tab:mask-bench} reports the measurements. Masking the decryption
product roughly doubles that routine---a factor of $1.95$--$1.98$ across all
parameter sets---matching the ``two multiplications instead of one'' prediction
of Section~\ref{sec:masking}. On the full decapsulation the overhead is only
$1.32$--$1.33\times$, because decapsulation also re-encrypts (the
Fujisaki--Okamoto transform) and decodes, so the extra multiplication is a
smaller share of the total. The two variants cost essentially the same time---the
XOR refresh and the DRBG draw are negligible next to the extra
multiplication---so the choice between them is dictated by security and key size,
not by speed. The stored-share variant enlarges the secret key by
$2\lceil n/8 \rceil - 32$ bytes, a factor of $2.89$--$2.96$, while the public key
and ciphertext are unchanged\footnote{The code with this countermeasure is available
 in \url{https://gitlab.inria.fr/eclair/hqcload}.}.

\begin{table}[t]
\centering
\caption{Cost of additive masking on the reference HQC implementation
(13th~Gen Intel Core i7-1365U, GCC $16.1.1$ and with \texttt{-O3}, median cycles). ``decrypt'' is
the isolated \texttt{hqc\_pke\_decrypt}; ``decaps'' is the full
\texttt{crypto\_kem\_dec}; $|\mathrm{sk}|$ is the secret-key size in bytes.
Factors are relative to the unmasked reference of each parameter set.}
\label{tab:mask-bench}
\small
\begin{tabular}{llrrrrrr}
\toprule
 & & \multicolumn{2}{c}{decrypt} & \multicolumn{2}{c}{decaps} & \multicolumn{2}{c}{$|\mathrm{sk}|$}\\
\cmidrule(lr){3-4}\cmidrule(lr){5-6}\cmidrule(lr){7-8}
set & mode & cycles & $\times$ & cycles & $\times$ & bytes & $\times$\\
\midrule
\multirow{3}{*}{HQC-1}
 & reference       & $2\,617\,290$  & $1.00$ & $7\,712\,648$   & $1.00$ & $2321$  & $1.00$\\
 & per-call        & $5\,102\,640$  & $1.95$ & $10\,190\,324$  & $1.32$ & $2321$  & $1.00$\\
 & stored+refresh  & $5\,091\,156$  & $1.95$ & $10\,184\,322$  & $1.32$ & $6707$  & $2.89$\\
\midrule
\multirow{3}{*}{HQC-3}
 & reference       & $7\,744\,522$  & $1.00$ & $23\,087\,538$  & $1.00$ & $4602$  & $1.00$\\
 & per-call        & $15\,270\,688$ & $1.97$ & $30\,638\,202$  & $1.33$ & $4602$  & $1.00$\\
 & stored+refresh  & $15\,227\,802$ & $1.97$ & $30\,524\,690$  & $1.32$ & $13534$ & $2.94$\\
\midrule
\multirow{3}{*}{HQC-5}
 & reference       & $18\,841\,256$ & $1.00$ & $56\,158\,166$  & $1.00$ & $7333$  & $1.00$\\
 & per-call        & $37\,242\,114$ & $1.98$ & $74\,492\,206$  & $1.33$ & $7333$  & $1.00$\\
 & stored+refresh  & $37\,146\,158$ & $1.97$ & $74\,357\,750$  & $1.32$ & $21711$ & $2.96$\\
\bottomrule
\end{tabular}
\end{table}

\subsection{Storing $y$ in the FAFFT transform domain}
\label{sec:transform}

Polynomial multiplication is a bottleneck for HQC performance,
and recently~\cite{ras2025} and~\cite{ChenCPY26}
have suggested accelerating this using the Frobenius Additive Fast
Fourier Transform (FAFFT) algorithm in place of the Karatsuba
multiplication proposed in the specification.
If this is done, then 
one can precompute and store the transform $\hat{y} := \fafft(y)$ as the
private-key representation, so that the sparse vector~$y$ is never loaded during
decapsulation.
This saves the forward FAFFT which would otherwise be applied to $y$ in
each decapsulation, to say nothing of the computation required to
recover $y$ from the secret \seed,
so it implies a significant time saving; however,
$\hat{y}$ has twice the length of $y$
(and far more than the length of \seed),
so this comes at the cost of a nontrivial space overhead.

The FAFFT of a (very) sparse vector is necessarily (very) dense
by the standard uncertainty principle for discrete Fourier transforms
(see e.g.~\cite{Donoho--Stark}).
However, given that $\omega$ is designed to be in $O(\sqrt{n})$,
the uncertainty principle actually only tells us that $\hw(\hat{y})$ is
in $\Omega(\sqrt{n})$.
In practice, though, the weight of $\hat{y}$ is variable,
and as a random sum of transform-domain basis vectors
we expect it to be generally large and distributed essentially as for
random vectors of length $2n$.
In particular,
The zero-word rate
of~$\hat{y}$ for random $y$ (of weight $\omega$)
is expected to be far below the $\approx 0.887$ our distinguisher relies on,
and observing the loads of~$\hat{y}$ therefore yields essentially no information.
Even if some zero words are detected, the weight of most nonzero words
should be relatively large---certainly beyond the limit where decoding
hints are useful.
Storing $\hat{y}$ instead of $y$ (or \seed)
should therefore provide some natural resistance to our attack.

The FAFFT is not required, or even suggested, by the current HQC specification,
so this
countermeasure is only available to newer implementations that already
use the FAFFT.
Moreover, generating the private key directly in the transform domain
(to protect \KEMKeyGen) is
non-trivial: the standard fixed-weight sampler works in the coefficient domain,
and converting the representation while preserving the public-key equation $s =
x + h \cdot y$ requires care. We leave a complete treatment, and a
detailed analysis of the impact on our attack, to future work.

\section{Conclusion}
We introduced a side-channel attack against HQC exploiting
leakage from the loading and storing of \ssvs during polynomial
multiplication. Rather than recovering secret bits directly, \emph{the attack
distinguishes zero from nonzero machine words and converts this information
into decoding hints.}
Reusing state-of-the-art decoding with hints~\cite{DAchilleEK26}, these hints
recover the secret key. For HQC-1, classifying at 32-bit granularity reveals up
to 88.7\% of the coordinates as zero, yielding a shortened syndrome-decoding
instance that can be solved much more efficiently than the original.

More broadly,
our results show that \emph{memory accesses alone can constitute a powerful
leakage source when manipulating highly sparse secrets}, highlighting the need
to consider compiler-generated code and load/store operations in side-channel
evaluations of PQC implementations.

\paragraph{Future work.}
More advanced distinguishers or template attacks may reduce the trace count
and improve leakage exploitation on more noisy targets.
Moreover, evaluating the attack across microarchitectures, compiler versions, and
optimization levels would clarify the impact of compiler-generated code on
leakage. Algorithmically, integrating the hints with soft-decision decoding,
and checking whether similar leakage affects other PQC schemes relying on
\ssvs, are natural next steps. Finally, the transform-domain countermeasure
calls for dedicated study: sampling fixed-weight vectors \emph{directly} in the
FAFFT domain---so the sparse coefficient-domain representation is never
materialised or loaded---while remaining consistent with $s = x + h\cdot y$ is,
to our knowledge, not well studied, and a complete implementation and security
evaluation remain open.

\section*{Acknowledgements}
We thank Pierre-Yvan Liardet, Rafael Carrera Rodriguez, Aurélien
Vasselle, and Guillaume Bethouart for useful discussions and help with equipment.

\bibliographystyle{alpha}
\bibliography{biblio}

\end{document}